\documentclass{pasj01}
\usepackage{url}

\Received{$\langle$reception date$\rangle$}
\Accepted{$\langle$acception date$\rangle$}
\Published{$\langle$publication date$\rangle$}

\RequirePackage{color}

\begin{document}

\title{The stellar halo of the Milky Way traced by blue horizontal-branch stars
in the Subaru Hyper Suprime-Cam Survey}

\author{Tetsuya~Fukushima\altaffilmark{1},
Masashi~Chiba\altaffilmark{1}, 
Mikito~Tanaka\altaffilmark{2}, 
Kohei~Hayashi\altaffilmark{3}, 
Daisuke~Homma\altaffilmark{4}, 
Sakurako~Okamoto\altaffilmark{6,4}, 
Yutaka~Komiyama\altaffilmark{4,5}, 
Masayuki~Tanaka\altaffilmark{4}, 
Nobuo~Arimoto\altaffilmark{7},
and Tadafumi~Matsuno\altaffilmark{4,5}
}

\altaffiltext{1}{Astronomical Institute, Tohoku University, Aoba-ku,
Sendai 980-8578, Japan \\E-mail: {\it t.fukushima@astr.tohoku.ac.jp}}
\altaffiltext{2}{Department of Advanced Sciences, Faculty of Science and Engineering,
Hosei University, 184-8584 Tokyo, Japan}
\altaffiltext{3}{ICRR, The University of Tokyo, Kashiwa, Chiba 277-8583, Japan}
\altaffiltext{4}{National Astronomical Observatory of Japan, 2-21-1 Osawa, Mitaka,
Tokyo 181-8588, Japan}
\altaffiltext{5}{The Graduate University for Advanced Studies, Osawa 2-21-1, Mitaka,
Tokyo 181-8588, Japan}
\altaffiltext{6}{Subaru Telescope, National Astronomical Observatory of Japan, 650 North A'ohoku Place,
Hilo, HI 96720, USA}
\altaffiltext{7}{Astronomy Program, Department of Physics and Astronomy,
Seoul National University, 599 Gwanak-ro, Gwanaku-gu, Seoul, 151-742, Korea}


\maketitle

\begin{abstract}
We report on the global structure of the Milky Way (MW) stellar halo up to its outer boundary 
based on the analysis of blue-horizontal branch stars (BHBs). These halo tracers are extracted
from the $(g,r,i,z)$ band multi-photometry in the internal data release of the on-going
Hyper Suprime-Cam Subaru Strategic Program (HSC-SSP) surveyed over $\sim550$~deg$^2$ area.
In order to select most likely BHBs by removing blue straggler stars (BSs) and other contamination
in a statistically significant manner, we have developed and applied an extensive Bayesian method,
instead of the simple color cuts adopted in our previous work, where each of the template BHBs
and non-BHBs obtained from the available catalogs is represented as a mixture of multiple Gaussian
distributions in the color-color diagrams.  We found from the candidate BHBs in the range of
$18.5<g<23.5$~mag that the radial density distribution over a Galactocentric radius
of $r=36-360$~kpc can be approximated as a single power-law profile with an index of
$\alpha=3.74^{+0.21}_{-0.22}$ or a broken power-law profile with an index of
$\alpha_{\rm in}=2.92^{+0.33}_{-0.33}$ at $r$ below a broken radius of
$r_{\rm b}=160^{+18}_{-19}$~kpc and a very steep slope of
$\alpha_{\rm out}=15.0^{+3.7}_{-4.5}$ at $r>r_{\rm b}$. The latter profile with
a prolate shape having an axial ratio of $q=1.72^{+0.44}_{-0.28}$ is most likely
and this halo may hold a rather sharp boundary at $r\simeq160$~kpc.
The slopes of the halo density profiles are compared with those from 
the suite of hydrodynamical simulations for the formation of stellar halos.
This comparison suggests that the MW stellar halo may consist of the two
overlapping components: the {\it in situ.} inner halo as probed by RR Lyrae stars showing
a relatively steep radial density profile and the {\it ex situ.} outer halo with
a shallow profile probed by BHBs here, which is made by accretion of small stellar systems.
\end{abstract}



\section{Introduction}

A stellar halo surrounding a disk galaxy like our Milky Way (MW) is thought to
have been developed through hierarchical assembly of small stellar systems such as
dwarf galaxies \citep{Searle1978}. Because of the long relaxation time in the halo,
the structure of a current stellar halo, including the distribution of both smooth
and non-smooth spatial features, reflects the past merging and accretion histories.
Indeed, many halo substructures have been identified in the form of
stellar streams in spatial coordinates as well as separate clumps
in phase space. The former substructures correspond to the merging events
within a few dynamical times, whereas the latter ones in phase space
persist over many billion years (e.g., \cite{Helmi1999,Bullock2005,Cooper2010}).

The smooth component of a stellar halo is also affected by the past merging
history. \citet{Deason2014} investigated the results of numerical simulation
for the merging-driven formation of a stellar halo by \citet{Bullock2005} and
showed that the slope of the density profile for the outer part of a stellar halo
depends on the average time of merging, in such a manner that the case of a more
recent merging time reveals a shallower radial density profile
over $50 < r/{\rm kpc} < 100$. It is also shown that the break in the stellar
halo slope, which might be present in the MW halo, can be made by tidal debris
from a merging satellite when it is at an apocenter position \citep{Deason2018b}.
Also, the recent suite of magneto-hydrodynamical numerical simulation for galaxy formation,
named Auriga \citep{Grand2017,Monachesi2018}, shows that both the slope in a density profile
of a simulated stellar halo and its metallicity gradient are intimately related to
the number of main progenitor satellites, which contribute to
the total mass of a final halo. It is thus of great importance to
derive the structure of a stellar halo to infer its merging history.

While the detection and analysis of stellar halos in external disk galaxies are challenging
because of their very faint brightness, the stars distributed in the MW halo
provide us with a unique opportunity to study the structure of the stellar halo
in great detail (see reviews, \cite{Helmi2008,Ivezic2012,Feltzing2013,Bland-Hawthorn2014}). 
The direct method probing the MW stellar halo is to use bright halo tracers including
red giant-branch (RGB) stars, RR Lyrae (RRL), blue horizontal-branch (BHB) stars
as well as blue straggler (BS) stars,
with which it is possible to map out the MW stellar halo out to its outer part
(e.g., \cite{Sluis1998,Yanny2000,Chen2001,Sirko2004,Newberg2005,Juric2008,Keller2008,
Sesar2011,Deason2011,Xue2011,Deason2014,Cohen2015,Cohen2017,Vivas2016,Slater2016,
Xu2018,Hernitschek2018}). These studies over a Galactocentric distance $r$ of a few tens kpc to
$\sim100$~kpc have revealed that the MW stellar halo includes a general smooth component,
which is often fit to a power-law density profile, and several irregular substructures
associated with recent merging events of dwarf galaxies, such as the Sagittarius
stream and Virgo overdensity \citep{Ibata1995,Belokurov2006,Juric2008}.

More recent studies have explored much distant halo regions beyond $r = 100$~kpc
to reach a possible virial radius of a MW-sized dark matter halo with $r \sim 300$~kpc and more
\citep{Hernitschek2018,Deason2018a,Fukushima2018,Thomas2018}.
This is because the outer parts of a stellar halo reflect the merging/accretion history
over past billion years \citep{Bullock2005,Deason2014,Pillepich2014,Monachesi2018}.
In particular, the outer boundary of the stellar halo may be present in the form of
a sharp outer edge or broadly extended without any clear cut depending on the recent
merging/accretion events. Among several halo tracers to probe the outskirts of the MW stellar
halo, BHB stars have been frequently adopted and analyzed in the large photometric surveys including
Subaru/Hyper Suprime-Cam (HSC) \citep{Deason2018a,Fukushima2018} and Canada-France Imaging Survey (CFIS)
\citep{Thomas2018}. \citet{Deason2018a} selected BHBs from the public data release of the
HSC Subaru Strategic Program (HSC-SSP) surveyed over $\sim 100$~deg$^2$ using
$griz$-band photometry and derived the power-law radial profile with an index $\alpha \simeq 4$.
Concurrently with the completion of this work, we elsewhere reported
\citep{Fukushima2018} our results using BHBs extracted from the internal data release
of HSC-SSP over $\sim 300$~deg$^2$. They derived a halo density profile between $r=50$~kpc and 300~kpc
and fit, after the subtraction of the fields containing known substructures, to either
a single power-law model with $\alpha \simeq 3.5$ and an axial ratio of $q \simeq 1.3$ or
a broken power-law model with an inner/outer slope of $3.2/5.3$ at a break radius of 210~kpc.
More recently, \citet{Thomas2018} presented their analysis of BHBs selected using deep $u$-band
imaging from the CFIS survey combined with $griz$-band data from Pan-STARRS~1. They show that
a broken power-law model with an inner/outer slope of $4.24/3.21$ at a break radius of
41.4 kpc is the best fitting case out to $r \sim 220$~kpc.

The main obstacle in the selection of BHBs from photometric data is to remove the contaminants
having similar colors and magnitudes to BHBs, such as BSs,
white dwarfs (WDs), QSOs, as well as distant faint galaxies having point-source images. 
This issue is more important in the outer parts of the halo, where the number density of BHBs
becomes quite sparse compared with the contaminants. In our previous work \citep{Fukushima2018},
we use the HSC-SSP data obtained until 2016 April (internal data release S16A) and select BHBs
located inside specific regions in the color-color diagrams defined in the combination of
$griz$ band. This selection method of BHBs based on the simple color cuts provides
basically the same results as those based on the maximum likelihood method, where
the probability distribution of each stellar population is given as a single Gaussian
in $griz$ space (see also \cite{Deason2018a}). The current paper is an extension of our previous
work, in which we use the most recent internal data release of HSC-SSP covering $\sim 550$~deg$^2$
and develop an extensive Bayesian method to minimize the effects of non-BHB contamination as much
as possible. We also consider the distribution of BS stars to obtain the additional information
on the structure of the MW stellar halo.

This paper is organized as follows.
In Section 2, we present the data that we utilize here and the method for the selection
of candidate BHBs based on the $griz$-band photometric data obtained in the HSC-SSP survey.
Our Bayesian method for the selection of BHB stars and their spatial distribution is also
described. In Section 3, we show the results and discussion of our Bayesian analysis
for the best set of parameters of the spatial distribution of BHB stars.
Our conclusions are drawn in Section 4.

\begin{table}
\tbl{Obseved Regions with HSC-SSP}{%
\begin{tabular}{lcccccc}
\hline
Region &   RA  &  DEC  &  $l$  &  $b$  & Adopted area & Use    \\
       & (deg) & (deg) & (deg) & (deg) &  (deg$^2$)   & Yes/No \\
\hline \hline
XMM-LSS   &   35  & $-5$ & 170 & $-59$ & 60  & No\\
WIDE12H   &  180  &    0 & 276 &   60  & 68  & Yes\\
WIDE01H   &   19  &    0 & 136 & $-62$ &  0  & No\\
VVDS      &  337  &    0 &  65 & $-46$ & 169 & Yes\\
GAMA15H   &  217  &    0 & 347 &   54  & 85  & No\\
GAMA09H   &  135  &    0 & 228 &   28  & 92  & Yes\\
HECTOMAP  &  242  &   43 &  68 &   47  & 75  & Yes \\
\hline
AEGIS     &  214  &   51 &  95 &   60  & 2.5 & Yes\\
\hline
\end{tabular}}\label{tab: region}
\end{table}

\section{Data and Method}

\subsection{Data}

We make use of the imaging data obtained from the HSC-SSP Wide survey, which plans
to cover $\sim 1,400$ deg$^2$ in five photometric bands ($g$, $r$, $i$, $z$, and $y$)
\citep{Aihara2018a,Aihara2018b,Furusawa2018,Kawanomoto2018,Komiyama2018,Miyazaki2018}.
In this Wide layer, the target 5$\sigma$ point-source
limiting magnitudes are ($g$, $r$, $i$, $z$, $y$) = (26.5, 26.1, 25.9, 25.1, 24.4) mag.
In this work, we adopt the $g$, $r$, $i$ and $z$-band data obtained before 2018 April
(internal data release S18A), for the selection of BHBs and the removal of
other contaminants as explained below.
The data set covers six separate fields along the celestial equator, named XMM-LSS, 
WIDE12H, WIDE01H, VVDS, GAMA15H and GAMA09H,  a field named HECTOMAP around
$(\alpha_{\rm 2000},\delta_{\rm 2000})=(242^{\circ},43^{\circ})$, and 
a calibration field named AEGIS around $(240^{\circ},51^{\circ})$,
amounting to $\sim 550$ deg$^2$ in total (See Table \ref{tab: region}).
Since WIDE01H has no $i$ and $z$-band data, we do not use this region. 
The total area that the current data set covers is to be compared with
$\sim 300$ deg$^2$ covered in our previous analysis of BHBs from the data obtained
before 2016 April \citep{Fukushima2018}.
The HSC data are processed with
hscPipe v6.5 \citep{Bosch2018}, a branch of the Large Synoptic Survey Telescope pipeline
\citep{Ivezic2008,Axelrod2010,Juric2017} calibrated against PS1 DR1
photometry and astrometry \citep{Schlafly2012,Tonry2012,Magnier2013}.
All the photometry data are corrected for the mean Galactic foreground
extinction \citep{Schlafly2011}.

We note that as shown in \citet{Fukushima2018}, both GAMA15H and XMM-LSS contain
several spatial substructures associated with the Sagittarius (Sgr) stream,
which is formed from a tidally disrupting, polar-orbit satellite, Sgr dwarf.
Our interest in this paper is to deduce the structure of the smooth halo component,
thus we exclude these fields in the following analysis.

\subsection{Selection of targets}

\begin{figure}[t]
\includegraphics[width=80mm]{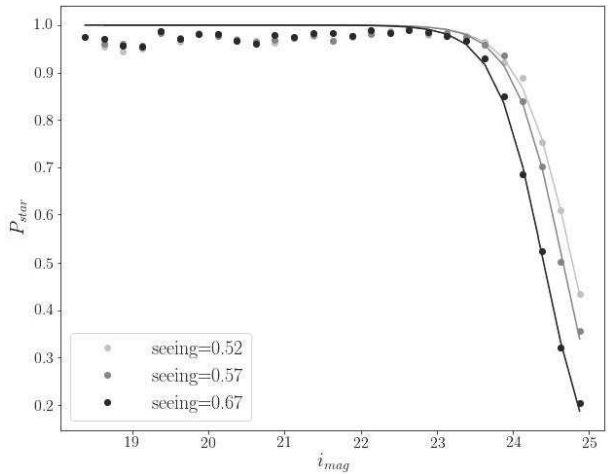}
\hspace*{10mm}
\caption{
The circles denote the fraction of stars classified as HST/ACS among HSC-classified stars
for three different seeings of $0''.67$, $0''.57$ and $0''.52$.
This fraction is close to one at $i<23$ and decreases to $\sim 0.5$
at $i=24.5$ at the high seeing of the survey ($0''.67$).
The lines show the fitted functions given in Equation (\ref{eq: Pstar}).
}
\label{fig: Pstar}
\end{figure}

For the analysis of BHBs from our current sample, we select point sources
using the {\it extendedness} parameter from the pipeline, namely
{\it extendedness}$=0$ for point sources and {\it extendedness}$=1$ for
extended images like galaxies. This parameter is computed based on
the ratio between PSF and cmodel fluxes \citep{Abazajian2004}, where a point source
is defined to be an object having this ratio larger than 0.985. As shown in \citet{Aihara2018b},
this star/galaxy classification becomes uncertain for faint sources.
The contamination, defined as the fraction of galaxies classified as HST/ACS
among HSC-classified stars, is close to zero at $i<23$, but increases to
$\sim 50\%$ at $i=24.5$ at the median seeing of the survey ($0.7$ arcsec).
These properties are summarized in Figure \ref{fig: Pstar}.
In what follows, we adopt point sources with $i \le 24.5$ and
investigate the possible effect of the contamination by faint galaxies.

We then select point sources in the following magnitude and color ranges:
\begin{eqnarray}
 18.5 < &g& < 23.5 \nonumber \\
 -0.3 < g&-&r < 0 \nonumber \\
-0.4< r&-&i <0.4 \nonumber \\ 
 -0.25< i&-&z <0.1  \ ,
\label{eq: sampleslection}
\end{eqnarray}
where the faint limit for the $g$-band magnitude range is taken based on its photometric error of
typically $\simeq 0.05$ mag with maximum of $\simeq 0.1$ mag

These point-source samples include not only BHBs but also other point sources including BSs,
WDs and QSOs, with some amount of faint galaxies which are missclassified as stars.
As demonstrated in \citet{Fukushima2018}, BHBs are distributed in the distinct region in the
$i-z$ vs. $g-r$ diagram, because the $i-z$ color is affected by the Paschen features of stellar spectra
and is sensitive to surface gravity \citep{Lenz1998,Vickers2012}. Thus, other A-type stars having
higher surface gravity, i.e. BSs, as well as WDs can be excluded based on their distributions
in the $i-z$ vs. $g-r$ diagram. Since QSOs are largely overlapping with BHBs in this diagram,
the removal of these point sources also requires the use of the $g-z$ vs. $g-r$ diagram.

In our previous work reported in \citet{Fukushima2018}, we defined the likely bounding regions in these
color-color diagrams based on the locations of candidate BHBs identified by SDSS ($u$-band selected BHBs
and those selected from spectroscopy) and then selected most likely BHBs from our sample, which are
located inside the corresponding color-color regions.
However, this method still accompanies some contaminants in the selected BHB sample,
because the boundaries in the color-color diagrams are determined arbitrarily.

This paper instead adopts a Bayesian method for the selection of BHB stars, given the likely 
distribution for each of BHBs, BSs, WDs, QSOs and faint galaxies in the color-color diagrams
defined by $g$, $r$, $i$ and $z$-band. 

\begin{figure*}[t!]
\begin{center}
\includegraphics[width=180mm]{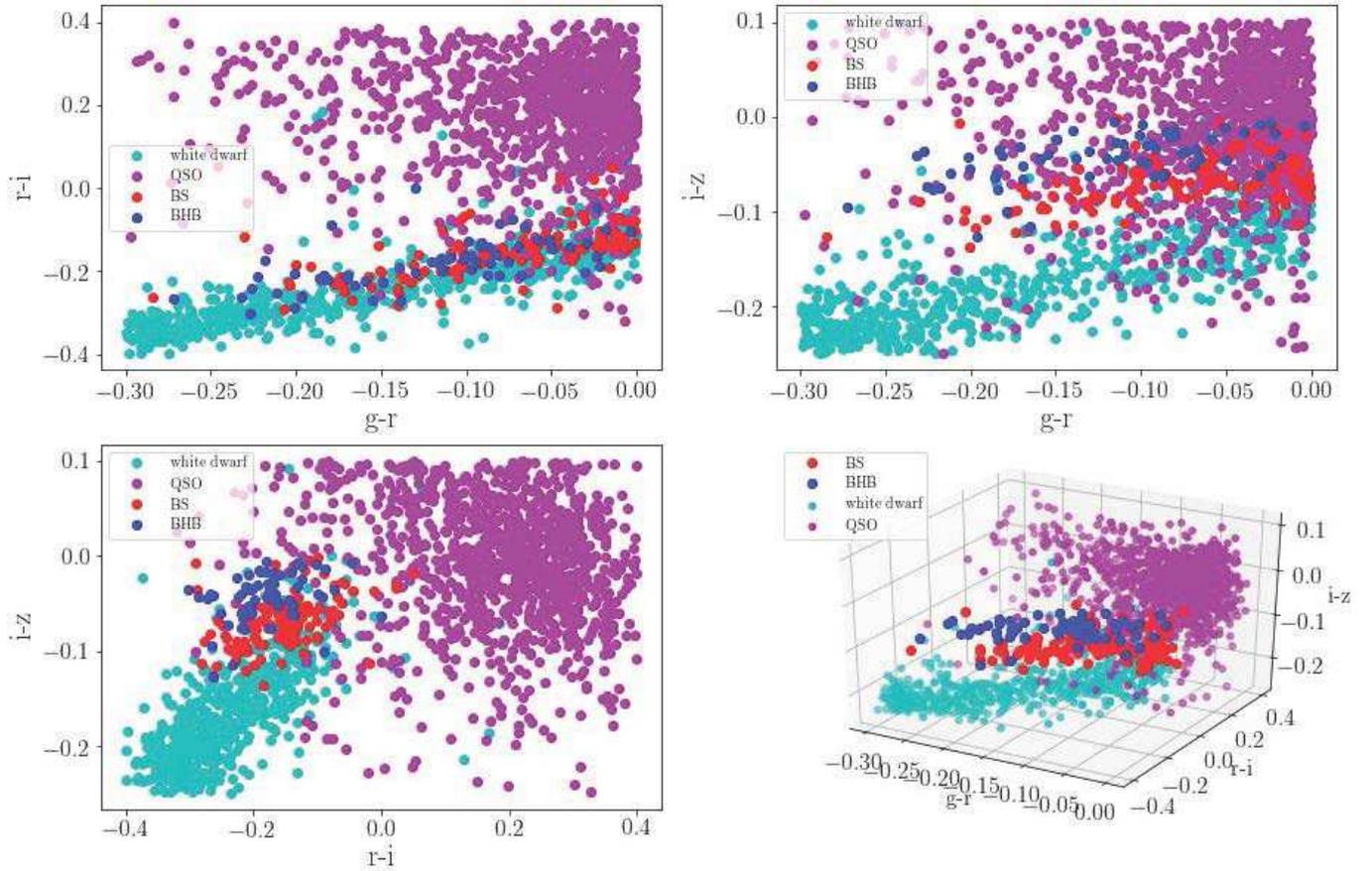}
\end{center}
\hspace*{10mm}
\caption{
The color-color diagrams for each of objects, WDs (cyan), QSOs (magenta), BSs (red) and
BHBs (blue circles) in the $g-r$ vs. $r-i$ space (upper left panel), the $g-r$ vs. $i-z$ space
(upper right panel) and the $r-i$ vs. $i-z$ space (lower left panel).
The lower right panel shows the three dimensional diagram in the $g-r$, $r-i$ and $i-z$ colors.
It follows that we can distinguish these objects in these color-color diagrams.
}
\label{fig: color-color}
\end{figure*}

\subsection{Probability distributions of BHBs, BSs, WDs, QSOs and galaxies in the color-color diagrams}

In order to derive the likely probability distributions of BHBs, BSs, WDs, QSOs and galaxies
in the color-color diagrams defined by $g$, $r$, $i$ and $z$-band, we first construct the representative
sample for each of these objects by crossmatching the HSC-SSP data with the corresponding data set
taken from several other works. The result is summarized in Figure \ref{fig: color-color}.

For WDs, we adopt the catalog taken from \citet{Kleinman2013,Kepler2015,Kepler2016},
which is selected from SDSS spectroscopy, and crossmatch with the current HSC-SSP data, resulting
in 596 WDs (cyan in Figure \ref{fig: color-color}).
For QSOs, we use the work by \citet{Paris2018}\footnote{\url{http://www.sdss.org/dr14/algorithms/qso_catalog}},
which contains 526,356 quasars from SDSS in the redshift range of $0.9 < z < 2.2$.
After crossmatching with HSC-SSP, we obtain 1055 QSOs (magenta in Figure \ref{fig: color-color}).

\begin{figure}[t!]
\begin{center}
\includegraphics[width=80mm]{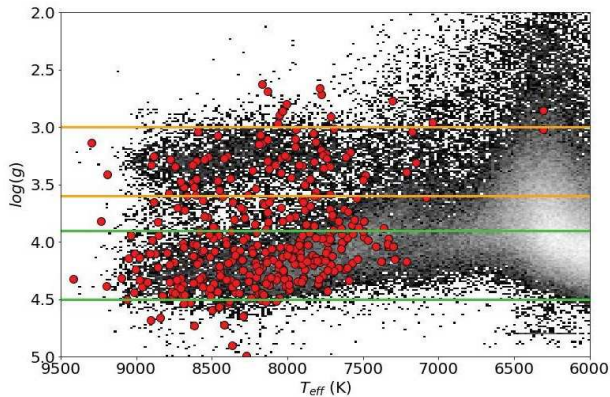}
\end{center}
\hspace*{10mm}
\caption{
The relation between effective temperatures, $T_{\rm eff}$, and surface gravities, $\log (g)$,
for the stars in SDSS/SEGUE DR15, which are shown with their densities in each bin (black shaded
squares) such that less dense bin is drawn with thicker black.
Among these sample stars, those crossmatched with the HSC-SSP data are shown as filled red circles.
The adopt ranges of $\log g$ for separating BHBs and
BSs are given as yellow ($3.0< \log (g) <3.6$) and green lines ($3.9 < \log (g) < 4.5$).
}
\label{fig: tefflogg}
\end{figure}

For BHBs and BSs, in contrast to our previous work \citep{Fukushima2018}, which adopted
the data in a dwarf spheroidal galaxy, Sextans, in the HSC-SSP footprint, we extract and select
the corresponding types of stars in the MW halo taken from SDSS DR15
\footnote{\url{http://skyserver.sdss.org/dr15/en/home.aspx}}
having the stellar atmospheric parameters provided from SEGUE (Sloan Extension for Galactic
Understanding and Exploration) Stellar Parameter Pipeline (SSPP: \cite{Lee2008}).
We set the constraints of $3.0 < \log (g) < 3.6$ for BHBs and $3.9 < \log (g) < 4.5$ for BSs,
which well separate the both stellar populations (Figure \ref{fig: tefflogg}).
We note that we set tighter constraints for this selection than those in \citet{Vickers2012},
which set $3.0< \log (g) <3.75$ for BHBs and $3.75 < \log (g) < 5.0$ for BSs, although
the final results remain basically unchanged.
The main reason to adopt the BHBs and BSs in the MW halo field, instead of Sextans,
to construct a template sample for the selection of these stars from HSC-SSP
is that there may exist systematic differences
in stellar ages and/or metallicities between the general halo field and Sextans.
To further remove possible systematics associated with the magnitude range of stars, which
originates from the age/metallicity difference between inner and outer halo components,
we crossmatch these SDSS data of the MW halo stars with the current HSC-SSP data and extract
the list of BHBs and BSs in the current sample,
which are depicted as filled blue circles in Figure \ref{fig: tefflogg}.

For galaxies as remaining contaminants, we use the HSC-SSP data with {\it extendedness}$=1$,
corresponding to extended images. 

Figure \ref{fig: color-color} shows the locations of BHBs, BSs, WDs and QSOs
in the color-color diagrams defined with $g$, $r$, $i$ and $z$-band. 
It follows that we can separate QSOs from other objects using $r-i$ color
and classify BHBs, BSs and WDs using $i-z$ color, as mentioned in the previous subsection. 

Next, to use these distributions of different objects in the color-color diagrams
for the application of a Bayesian method described below,
we construct the probability distribution function, $p(griz \mid {\rm Comp})$, for each population
(${\rm Comp}=$QSO, WD, BHB, BS and galaxy)
in terms of the mixture of several Gaussian distributions.
For this purpose. we use an extreme deconvolution Gaussian mixture model
(XDGMM\footnote[2]{\url{https://github.com/tholoien/XDGMM}}; \citet{Bovy2011} and \citet{Holoien2017}) 
with Python module, which allows us to estimate the best fit parameter for the given number of
Gaussian distributions and calculate Bayesian information criterion (BIC)
\footnote{Given the number of data points, $N$, the number of parameters, $k$, and the maximized
value of the likelihood function, $L_{\rm max}$, BIC is defined as
${\rm BIC} = k \ln (N) - L_{\rm max}$.} for each number.
We thus obtain the best fit parameter for each Gaussian given the lowest BIC. 

\begin{figure*}[t!]
\begin{center}
\includegraphics[width=180mm]{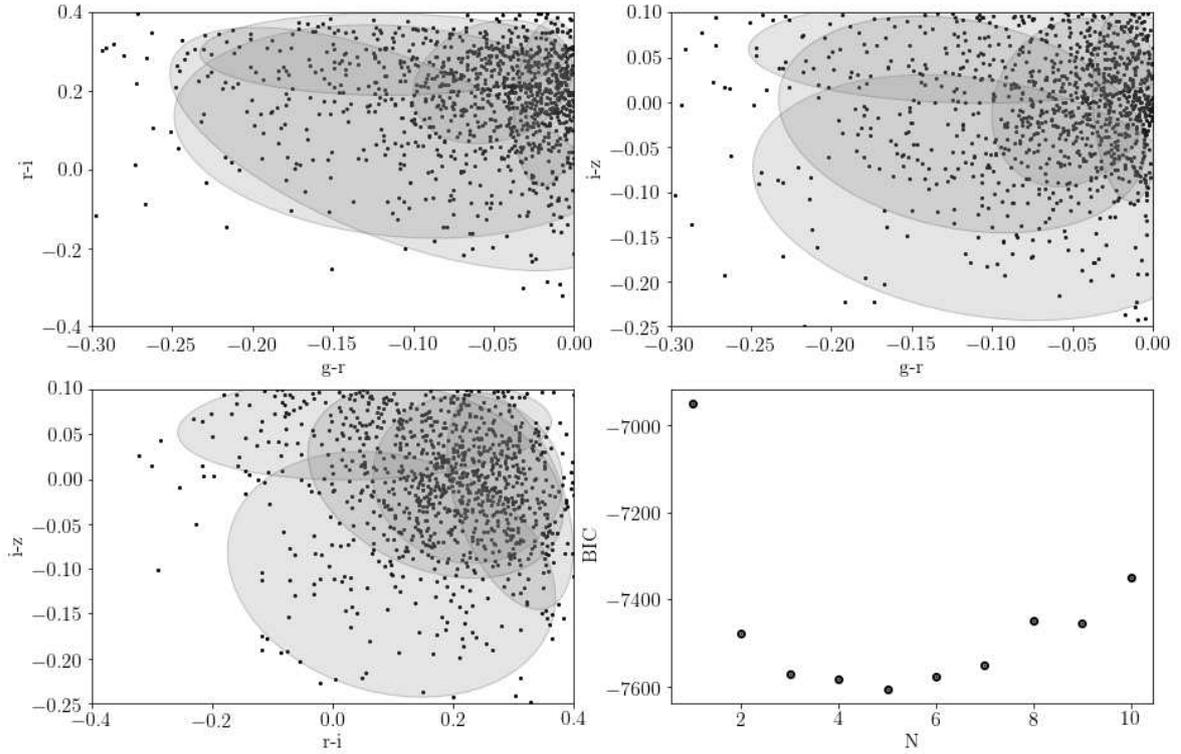}
\end{center}
\hspace*{10mm}
\caption{
An example of applying XDGMM for the case for QSOs to represent their distributions
in the color-color diagrams as the mixture of five Gaussian distributions.
The black points show the crossmatched QSOs with the HSC-SSP and gray shaded regions
demonstrate each Gaussian distribution. The lower right panel shows BIC as a function of
the number of Gaussian distributions, which suggests that the 5-componet model
reproduces the data most precisely.  
}
\label{fig: XDGMM}
\end{figure*}

For example, to obtain the probability $p(griz \mid {\rm QSO})$ of QSOs,
we provide one to ten Gaussian distributions and
adopt the case giving the lowest BIC. Figure \ref{fig: XDGMM} shows this result
for QSOs and the pdf can be reproduced by five Gaussian distributions.
Our experiments lead to $N_{{\rm Comp}}=$ 4, 5, 2, 1 and 9 for WDs, QSOs, BSs, BHBs
and galaxies, respectively. This is given as
\begin{eqnarray}
p(griz \mid {\rm Comp}) &=& \sum_{N_{{\rm Comp}}} G_{{\rm Comp}}(griz) 
\label{eq: pdf}
\end{eqnarray}
where `Comp' denotes each population (QSO, WD, BHB, BS and galaxy) and
$G(griz)$ is a three-dimensional normal distribution in $g-r$, $r-i$, and $i-z$ 
which is estimated using XDGMM.

\subsection{Contamination of galaxies}

As mentioned above (Figure \ref{fig: Pstar}), in our point-source sample selected with
{\it extendedness}$=0$, there still exist some amount of faint galaxies as contaminants
at the faint magnitude range of $i > 23$,  because of the difficulty for faint sources
to perform star/galaxy separation. To consider this contamination effect of galaxies in
the following analysis, we adopt the classification accuracy as a function of
the $i$-band magnitude and $i$-band seeing shown in Figure \ref{fig: Pstar}. The accuracy
is calculated by the fraction of stars classified as HST/ACS among HSC-classified stars
and we fit this fraction with the following function:
\begin{eqnarray}
P_{\rm star}(i) = \frac{1}{1 + \exp(ai + b)} \ ,  
\label{eq: Pstar}
\end{eqnarray}
where $i$ represent $i$-band magnitude and $(a, b)$ are the free parameters.

To take into account the effect of the seeing in $P_{\rm star}$, we obtain this function
for each of the three seeing cases of $0''.67$, $0''.57$ and $0''.52$.
In what follows, we adopt $P_{\rm star}$, for which the seeing is closest to
the one in the data we use here, ranging from $0''.545$ to $0''.62$.

\subsection{Distance estimates and spatial distributions for sample objects}

In addition to the probability distribution in the color-color diagrams,
we require the density distribution
for each population as functions of the $g$-band magnitude and spatial coordinates.

For both QSOs and galaxies, we assume, for simplicity, a constant density distribution
without depending on the $g$-band magnitude and spatial coordinates, although there may
exist some large scale structures.

For WDs, we adopt a disk-like spatial distribution given by \citet{Juric2008}, as also used by
\citet{Deason2014}, which assumes an exponential profile and has contributions
from thin and thick disk populations. Using the cylindrical coordinates $(R,z)$,
\begin{eqnarray}
\rho_{\rm thin} &=& \exp (R_0/L_1) \exp (-R/L_1-|z + z_0|/H_1) \nonumber \\
\rho_{\rm thick}&=& \exp (R_0/L_2) \exp (-R/L_2-|z + z_0|/H_2) \nonumber \\
\rho_{\rm disk} &=& \rho_{\rm thin} + \rho_{\rm thick} \ ,  
\label{eq: density_WD}
\end{eqnarray}
where $H_1=0.3$~kpc, $L_1=2.6$~kpc, $H_2=0.9$~kpc, $L_2=3.6$~kpc, $z_0=0.025$~kpc, $R_0=8.5$~kpc.
An absolute magnitude for WDs is taken from the model made by \citet{Deason2014}
with $\log (g_s) =8.0(7.5)$:
\begin{eqnarray}
M_g^{\rm WD} = 12.249 + 5.101(g-r),
\label{eq: Mg_WD}
\end{eqnarray}
where the error is given as $\sigma_{M_g^{\rm WD}} \simeq 0.5$ mag.

For the density distributions of BHBs and BSs,
we assume several models and estimate the associated parameters
using Goodman \& Weare's Affine Invariant Markov chain Monte Carlo (MCMC) \citep{Goodman2010},
which makes use of the Python module emcee\footnote[3]{\url{https://github.com/dfm/emcee}}
\citep{Foreman-Mackey2013} and judge these models based on BIC.
We note that both \citet{Deason2014} and \citet{Fukushima2018} adopt the same model parameters
for the spatial distributions of BHBs and BSs. However, this is not necessarily the case as
\citet{Thomas2018} demonstrated for several halo tracers of RRLs, BHBs and G dwarfs,
so we estimate the model parameters for BHBs and BSs separately.

In this study, we adopt the following five models:
\begin{itemize}
\item Spherical single power-law (SSPL)
\begin{eqnarray}
\rho_{{\rm halo}}(r) \propto r^{-\alpha},  \  r^2 = x^2+y^2+z^2  \ ,
\end{eqnarray}
where $\alpha$ denotes the power-law index for the stellar density distribution.

\item Spherical broken power-law (SBPL)
\begin{eqnarray}
\rho_{{\rm halo}}(r) \propto \left\{ 
\begin{array}{l} r^{-\alpha_{{\rm in}}} \ \ \  r \leq r_b \\ r^{-\alpha_{{\rm out}}} \ \ \  r>r_b \ ,
\end{array} \right. 
\end{eqnarray}
where $\alpha_{\rm in}$ and $\alpha_{\rm out}$ denote the power-law indices
in inner and outer halo regions, respectively, divided at the broken radius, $r_b$.

\item Axially symmetric single power-law (ASPL)
\begin{eqnarray}
\rho_{{\rm halo}}(r_q) \propto r_q^{-\alpha},
   \begin{array}{r} r_q^2 = x^2+y^2+z^2q^{-2} \ , \end{array}
\end{eqnarray}
where $q$ denotes the axis ratio.

\item Axially symmetric broken power-law (ABPL)
\begin{eqnarray}
\rho_{{\rm halo}}(r_q) \propto \left\{ 
\begin{array}{l} r_q^{-\alpha_{{\rm in}}} \ \ \  r_q \leq r_b \\ r_q^{-\alpha_{{\rm out}}} \ \ \  r_q>r_b 
\end{array} \right.
\end{eqnarray}

\item The Einast profile \citep{Einasto1965}
\begin{eqnarray}
\rho_{{\rm halo}}(r_q) \propto \exp [ -d_n((r_q/r_{{\rm eff}})^{1/n}-1)] \ ,
\end{eqnarray}
where $d_n = 3n-0.3333+0.0079/n$ for $n \geq 0.5$ \citep{Graham2006}.
This density profile is determined by $n$ and $r_{\rm eff}$, where
for larger (smaller) $n$, the inner profile at $r_q < r_{\rm eff}$ 
is steeper (shallower) than the outer one at $r_q > r_{\rm eff}$.

\end{itemize}

To obtain distance estimates for BHBs, we adopt the formula for their $g$-band absolute magnitudes,
$M_g^{\rm BHB}$, calibrated by \citet{Deason2011},
\begin{eqnarray}
M_g^{\rm BHB} &=& 0.434 - 0.169(g_{\rm SDSS}-r_{\rm SDSS})  \nonumber \\
      & &+ 2.319(g_{\rm SDSS}-r_{\rm SDSS})^2  + 20.449(g_{\rm SDSS}-r_{\rm SDSS})^3  \nonumber \\
     & &+ 94.517(g_{\rm SDSS}-r_{\rm SDSS})^4  ,
\label{eq: Mg_BHB}
\end{eqnarray}
where both $g$ and $r$-band magnitudes are corrected for interstellar absorption.
To estimate the absolute magnitude of BHBs selected from the HSC-SSP data, we use
Equations (\ref{eq: conversion-g}) - (\ref{eq: conversion}) below to translate HSC to SDSS filter system.
We then estimate the heliocentric distances and the three dimensional positions of BHBs
in rectangular coordinates, $(x,y,z)$, for the Milky Way space, where the Sun is
assumed to be at (8.5,0,0)~kpc. To consider the finite effect of contamination from BS stars
as shown below, we adopt their $g$-band absolute magnitudes, $M_g^{\rm BS}$, given by \citet{Deason2011},
\begin{equation}
M_g^{\rm BS} = 3.108 + 5.495 (g_{\rm SDSS}-r_{\rm SDSS}) .
\label{eq: Mg_BS}
\end{equation}
where the typical error is $\sigma _{M_g^{\rm BS}} \simeq 0.5$.

To estimate their absolute magnitudes, we convert the current HSC filter system to
the SDSS one by the formula given as \citet{Homma2016}
\begin{eqnarray}
g_{\rm HSC} &=& g_{\rm SDSS} - a (g_{\rm SDSS} - r_{\rm SDSS}) - b   \label{eq: conversion-g} \\
r_{\rm HSC} &=& r_{\rm SDSS} - c (r_{\rm SDSS} - i_{\rm SDSS}) - d   \\
i_{\rm HSC} &=& i_{\rm SDSS} - e (r_{\rm SDSS} - i_{\rm SDSS}) + f   \\
z_{\rm HSC} &=& z_{\rm SDSS} + g (i_{\rm SDSS} - z_{\rm SDSS}) - h    ,
\label{eq: conversion}
\end{eqnarray}
where $(a,b,c,d,e,f,g,h) = (0.074, 0.011, 0.004, 0.001, 0.106, 0.003, 0.006, 0.006)$ and
the subscript HSC and SDSS denote the HSC and SDSS system, respectively.
These formula have been calibrated from both filter curves and spectral atlas
of stars \citep{Gunn1983}.

\begin{table*}
\tbl{Prior distribution for model parameters}{%
\begin{tabular}{l|l|l|ccc}
\hline
Model &   BHB                    & BS              &
          $f_{\rm BHB}$ & $f_{\rm WD}$ & $f_{\rm QSO}$      \\
\hline\hline
SSPL  &   $\alpha=$2-10          & $\alpha=$2-10   &
          0-1           &  0-1         &  0-1               \\
\hline
SBPL  &   $\alpha_{\rm in}=$2-10, $\alpha_{\rm out}=$2-10  &   $\alpha_{\rm in}=$2-10, $\alpha_{\rm out}=$2-10  &
          0-1           &  0-1         &  0-1               \\
      &   $r_{\rm b}/{\rm kpc}=$50-400                     &   $r_{\rm b}/{\rm kpc}=$50-400                     &
                        &              &                    \\
\hline
ASPL  &   $\alpha=$2-10, $q=$0.1-3     & $\alpha=$2-10, $q=$0.1-3   &
          0-1           &  0-1         &  0-1               \\
\hline
ABPL  &   $\alpha_{\rm in}=$2-10, $\alpha_{\rm out}=$2-10  &   $\alpha_{\rm in}=$2-10, $\alpha_{\rm out}=$2-10  &
          0-1           &  0-1         &  0-1              \\
      &   $r_{\rm b}/{\rm kpc}=$50-400, $q=$0.1-3          &   $r_{\rm b}/{\rm kpc}=$50-400, $q=$0.1-3          &
                        &              &                    \\
\hline
Einasto & $n=$0.1-100, $r_{\rm eff}/{\rm kpc}=$0.1-500     &  $n=$0.1-100, $r_{\rm eff}/{\rm kpc}=$0.1-500        &
          0-1           &  0-1         &  0-1              \\
      &   $q=$0.1-3                                        &   $q=$0.1-3          &
                        &              &                    \\
\hline
\end{tabular} }
\label{tab:prior_distribution}
\end{table*}

\begin{table*}
\tbl{Best fit parameters}{%
\begin{tabular}{l|l|l|cccc}
\hline
Model &   BHB                    & BS              &
          $f_{\rm BHB}$ & $f_{\rm WD}$ & $f_{\rm QSO}$ & $\Delta$BIC      \\
\hline\hline
SSPL  &   $\alpha=3.76^{+0.24}_{-2.20}$ & $\alpha=4.59^{+0.17}_{-0.17}$   &
          $0.200^{+0.036}_{-0.032}$ & $0.870^{+0.007}_{-0.008}$ & $0.249^{+0.006}_{-0.007}$ & 109  \\
\hline
SBPL  &   $\alpha_{\rm in}=2.78^{+0.35}_{-0.32}$, $\alpha_{\rm out}=13.7^{+4.1}_{-4.9}$
      &   $\alpha_{\rm in}=4.42^{+0.25}_{-0.22}$, $\alpha_{\rm out}=12.2^{+5.1}_{-4.4}$   &
          $0.218^{+0.031}_{-0.035}$ & $0.867^{+0.008}_{-0.007}$ & $0.248^{+0.008}_{-0.008}$  & 70 \\
      &   $r_{\rm b}/{\rm kpc}=199^{+17}_{-34}$   &  $r_{\rm b}/{\rm kpc}=82.7^{+22.0}_{-11.4}$  &
                        &              &                 &   \\
\hline
ASPL  &   $\alpha=3.74^{+0.21}_{-0.22}$, $q=1.87^{+0.61}_{-0.38}$
      &   $\alpha=4.42^{+0.18}_{-0.16}$, $q=1.45^{+0.17}_{-0.14}$     &
          $0.199^{+0.030}_{-0.030}$ & $0.865^{+0.007}_{-0.007}$ & $0.248^{+0.006}_{-0.006}$ & 54  \\
\hline
ABPL  &   $\alpha_{\rm in}=2.92^{+0.33}_{-0.33}$, $\alpha_{\rm out}=15.0^{+3.7}_{-4.5}$
      &   $\alpha_{\rm in}=4.14^{+0.22}_{-0.23}$, $\alpha_{\rm out}=15.5^{+3.1}_{-4.9}$  &
          $0.213^{+0.030}_{-0.029}$ & $0.864^{+0.006}_{-0.007}$ & $0.249^{+0.008}_{-0.008}$  & 0  \\
      &   $r_{\rm b}/{\rm kpc}=160^{+18}_{-19}$, $q=1.72^{+0.44}_{-0.28}$  
      &   $r_{\rm b}/{\rm kpc}=66.8^{+12.2}_{-7.6}$,   $q=1.43^{+0.17}_{-0.12}$          &
                        &              &                &    \\
\hline
Einasto & $n=1.23^{+1.00}_{-0.42}$, $r_{\rm eff}/{\rm kpc}=57.2^{+10.5}_{-14.0}$
        & $n=5.51^{+3.02}_{-1.88}$, $r_{\rm eff}/{\rm kpc}=3.35^{+3.98}_{-2.27}$     &
          $0.203^{+0.033}_{-0.029}$ & $0.864^{+0.007}_{-0.008}$ & $0.248^{+0.006}_{-0.006}$  & 24 \\
        & $q=1.91^{+0.48}_{-0.34}$  & $q=1.49^{+0.19}_{-0.12}$   &
                        &              &               &     \\
\hline
\end{tabular} }
\label{tab:best_fit}
\end{table*}

\begin{figure*}[t!]
\begin{center}
\includegraphics[width=80mm]{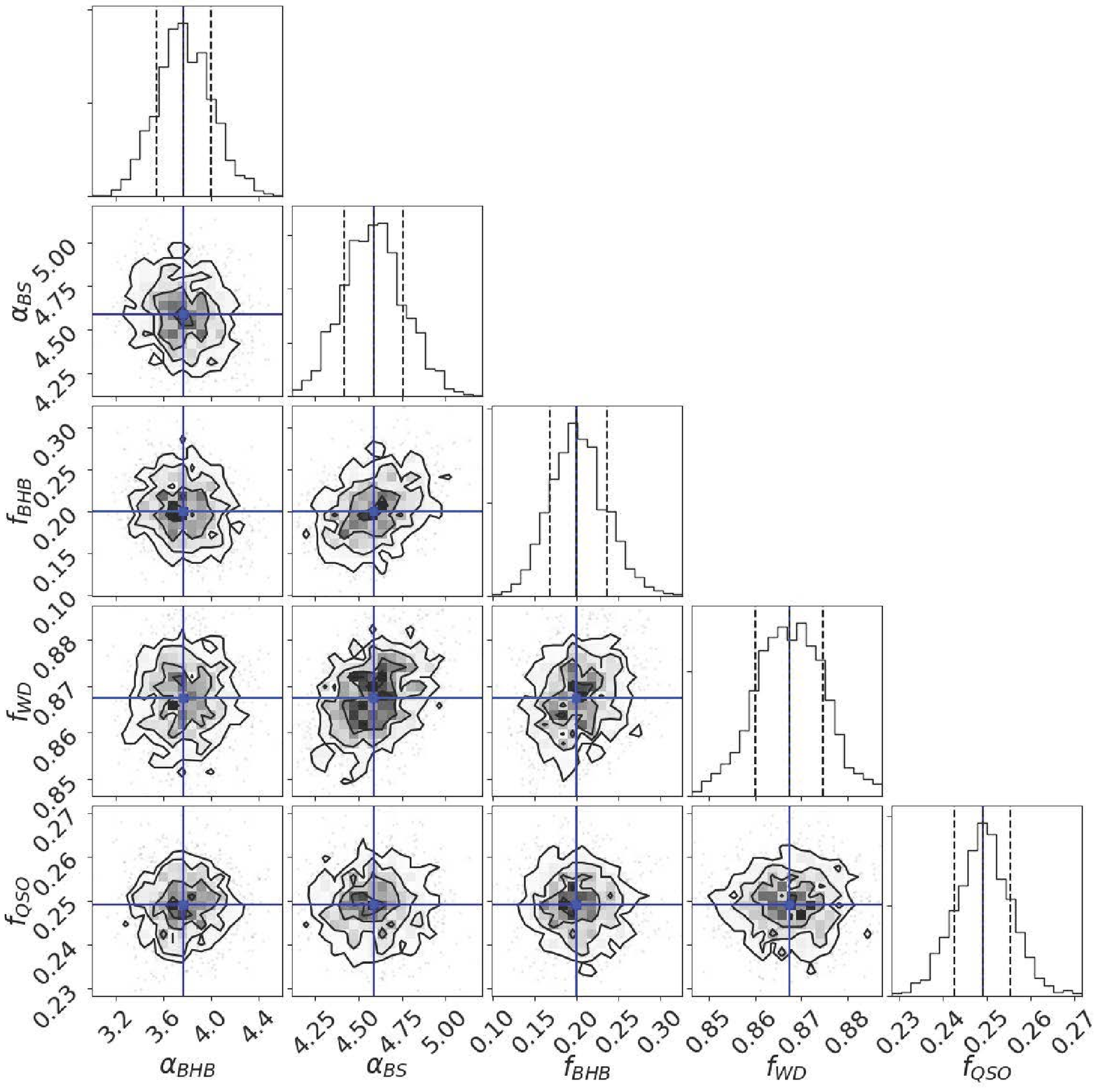}
\includegraphics[width=80mm]{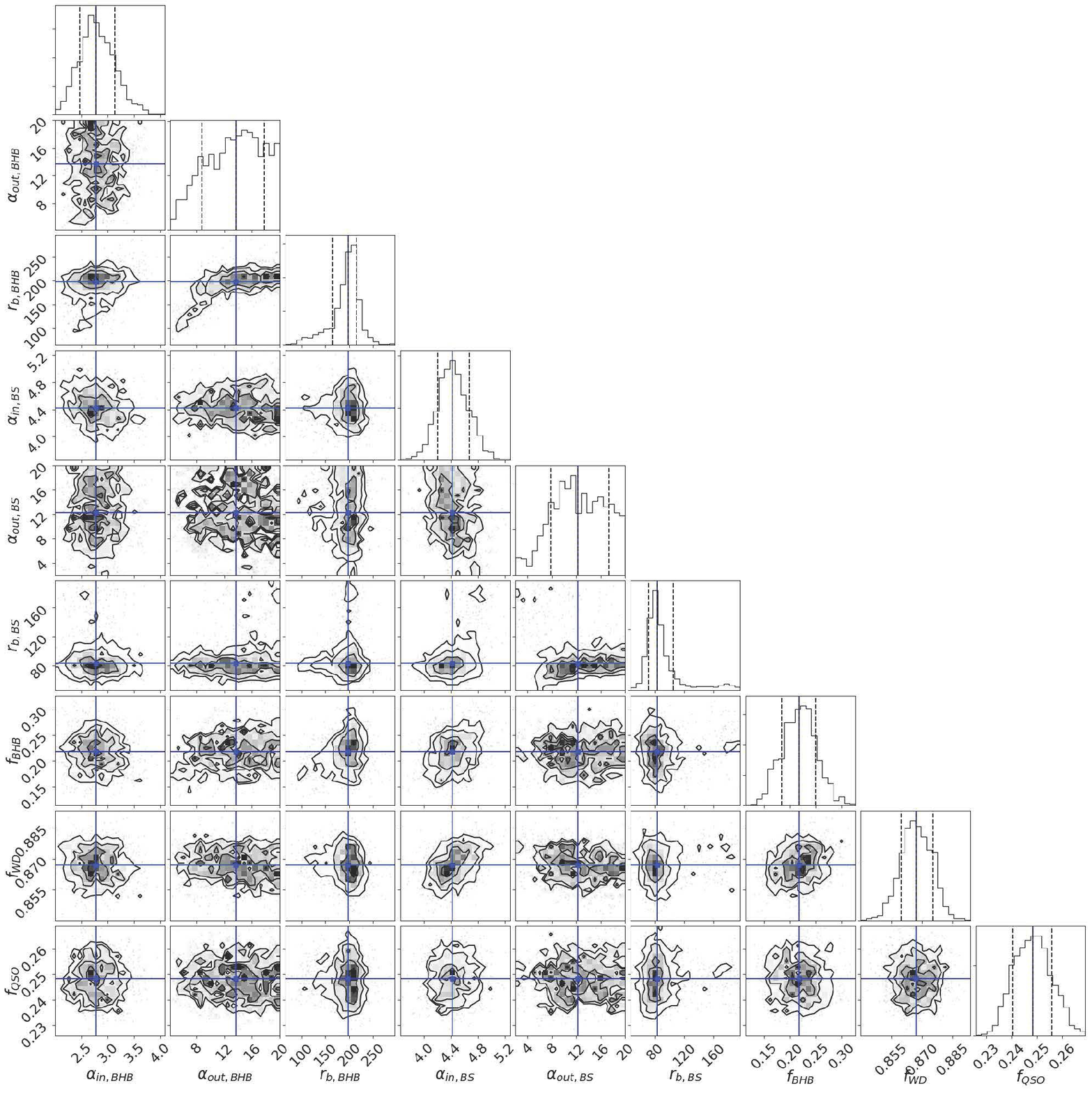}
\end{center}
\hspace*{10mm}
\caption{
MCMC results for SSPL (left) and SBPL (right panel).
}
\label{fig: Spherical}
\end{figure*}

\subsection{Maximum likelihood method for getting the radial density profile}

We maximize the likelihood defined as
\begin{eqnarray}
\ln \mathcal{L} = \sum_{i=1}^{N_{{\rm S}}} &\Bigl[  \Bigr. & \tilde{f}_{{\rm BHB}}
\tilde{\lambda}_{{\rm BHB}}(m_i, l_i, b_i, griz_i, seeing, \vec{\mu}_{\rm BHB})  \nonumber \\
& & + \tilde{f}_{{\rm BS}}\tilde{\lambda}_{{\rm BS}}(m_i, l_i, b_i, griz_i, seeing, \vec{\mu}_{\rm BS}) \nonumber \\
& & + \tilde{f}_{{\rm WD}}\tilde{\lambda}_{{\rm WD}}(m_i, l_i, b_i, griz_i, seeing)  \nonumber \\
& &+ f_{{\rm QSO}}\tilde{\lambda}_{{\rm QSO}}(m_i, l_i, b_i, griz_i, seeing)   \nonumber \\
 &+&  \tilde{\lambda}_{{\rm galaxy}}(m_i, l_i, b_i, griz_i, seeing)  \Bigl. \Bigr] \ ,
\end{eqnarray}
where the subscript $i$ denotes each object and the summation is performed over all the sample.
The fraction of each population ($\tilde{f}_{\rm WD}$, $\tilde{f}_{\rm BS}$, $\tilde{f}_{\rm BHB}$)
is defined by the following equations with four free parameters
($f_{\rm BHB}, f_{\rm BS}, f_{\rm WD}, f_{\rm QSO}$):
\begin{eqnarray}
& &\tilde{f}_{{\rm WD}} = f_{{\rm WD}}(1-f_{{\rm QSO}})  \\
& &\tilde{f}_{{\rm BS}} = (1-f_{{\rm BHB}})(1-f_{{\rm WD}})(1-f_{{\rm QSO}})  \\
& &\tilde{f}_{{\rm BHB}} = f_{{\rm BHB}}(1-f_{{\rm WD}})(1-f_{{\rm QSO}}) \ .
\end{eqnarray}
The function, $\tilde{\lambda}_{\rm Comp}$ with ${\rm Comp}=$ BHB, BS, WD, QSO and galaxy,
denotes the probability of each population having $m$ ($g$-band apparent magnitude),
Galactic coordiantes $(l, b)$, colors in $griz$, and the set of model parameters, $\vec{\mu}$,
given for the halo density distributions of BHBs and BSs (such as a power-law index and broken radius)
as introduced in the previous subsection.
This is given as
\begin{eqnarray}
& &\tilde{\lambda}_{{\rm Comp}}(m, l, b, griz, seeing, \vec{\mu}) = \nonumber \\
& &\int \Bigl[ \Bigr. G(m, griz, M) \lambda_{{\rm Comp}}(m, l, b, griz, seeing, \vec{\mu})  \Bigl. \Bigr]
 {\rm d}m \ {\rm d}(griz)  \ {\rm d}M \nonumber \\
& &  \Bigl/ \Bigr. \int \Bigl[  \lambda_{{\rm Comp}}(m, l, b, griz, seeing, \vec{\mu}) \Bigr] {\rm d}m \ {\rm d}(griz) \ {\rm d}l \ {\rm d}b 
\end{eqnarray}
where the denominator is a normalization over the ranges of $griz$, $m_g$, $l$ and $b$ 
specified in Equation (\ref{eq: sampleslection}) and the numerator is to consider photometric error and
deviation of absolute magnitude. 
$G(m, griz, M)$ is a fifth-dimensional normal distribution in $g-r$, $r-i$, $i-z$, apparent magnitude $m$
and absolute magnitude $M$, both in $g$-band in this work, i.e., $m_g$ and $M_g$.
Here, for simplicity, we assume that the functional dependence on each variable is separable,
so $G(m, griz, M)$ can be described as the multiplication of five one-dimensional normal distributions.
Because of only small deviation in $M_g$ for BHB, their normal distribution can be approximated as
a Dirac Delta so the integration for $M_g$ can be neglected.  

For each population with the color distribution $p(griz \mid {\rm Comp})$ given
in Equation (\ref{eq: pdf}) and with an estimated distance, $D$, we obtain the following equation.
\begin{itemize}
\item BHB
\begin{eqnarray}
& &\lambda_{{\rm BHB}}(m, l, b, griz, seeing, \vec{\mu}_{{\rm BHB}}) = \nonumber \\
& &  P_{{\rm star}} (m, seeing) p( griz \mid {\rm BHB}) \nonumber \\
& &\rho_{{\rm halo}}(X, Y, Z \mid m, l, b, gr, \vec{\mu}_{{\rm BHB}})  D^3(m, gr \mid {{\rm BHB}}) \cos(b)  
\end{eqnarray}
\item BS
\begin{eqnarray}
& &\lambda_{{\rm BS}}(m, l, b, griz, seeing, \vec{\mu}_{{\rm BS}}) = \nonumber \\
& &  P_{{\rm star}} (m, seeing) p( griz \mid {\rm BS}) \nonumber \\
& &\rho_{{\rm halo}}(X, Y, Z \mid m, l, b, gr, \vec{\mu}_{{\rm BS}})  D^3(m, gr \mid {{\rm BS}}) \cos(b)  
\end{eqnarray}
\item WD
\begin{eqnarray}
& &\lambda_{{\rm WD}}(m, l, b, griz, seeing) = \nonumber \\
& &  P_{{\rm star}} (m, seeing) p( griz \mid {\rm WD}) \nonumber \\
& &\rho_{{\rm disk}}(X, Y, Z \mid m, l, b, gr, \vec{\mu})  D^3(m, gr \mid {{\rm WD}}) \cos(b)  
\end{eqnarray}
\item QSO
\begin{eqnarray}
& &\lambda_{{\rm QSO}}(m, l, b, griz, seeing) = \nonumber \\
& &  P_{{\rm star}} (m, seeing) p( griz \mid {\rm QSO})   
\end{eqnarray}
\item galaxy
\begin{eqnarray}
& &\lambda_{{\rm galaxy}}(m, l, b, griz, seeing) = \nonumber \\
& &  (1-P_{{\rm star}} (m, seeing)) p( griz \mid {\rm galaxy})   
\end{eqnarray}
\end{itemize}

As described above, we estimate the best fit parameters using MCMC.
We assume the prior distribution is uniform over a concerned range
(see Table \ref{tab:prior_distribution}). The best-fit parameters
have been estimated using the 50th percentile of the posterior
distributions and the 16th and 84th percentiles have been used to
estimate the 1-$\sigma$ uncertainties.

\section{Results}

In this section, we show our main results following the Bayesian method shown in 
Section 2 and compare with our previous work based on the different method for the 
selection of BHBs using the S16A data of HSC-SSP.

\subsection{Best fit models}

\begin{figure*}[t!]
\begin{center}
\includegraphics[width=80mm]{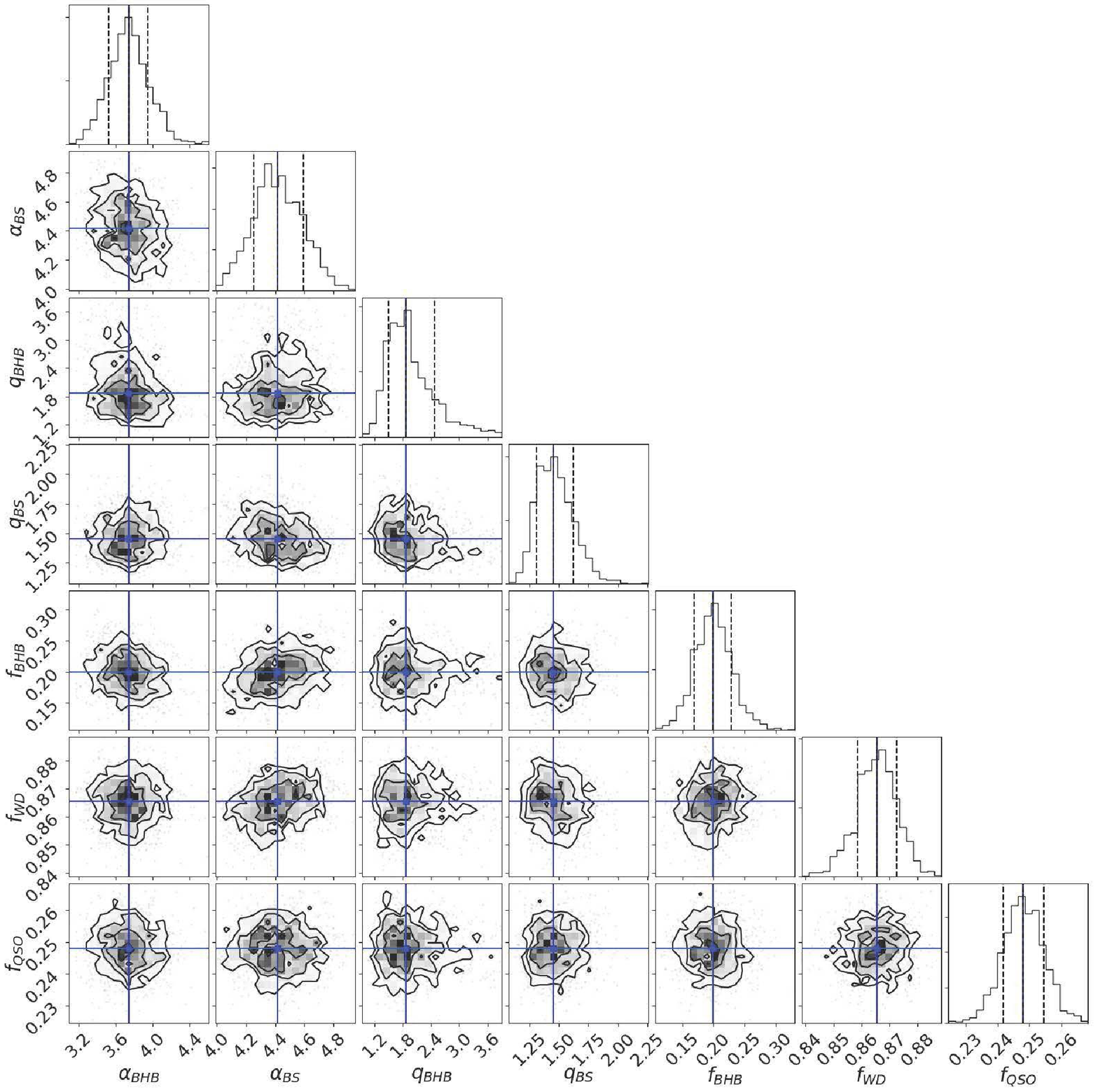}
\includegraphics[width=80mm]{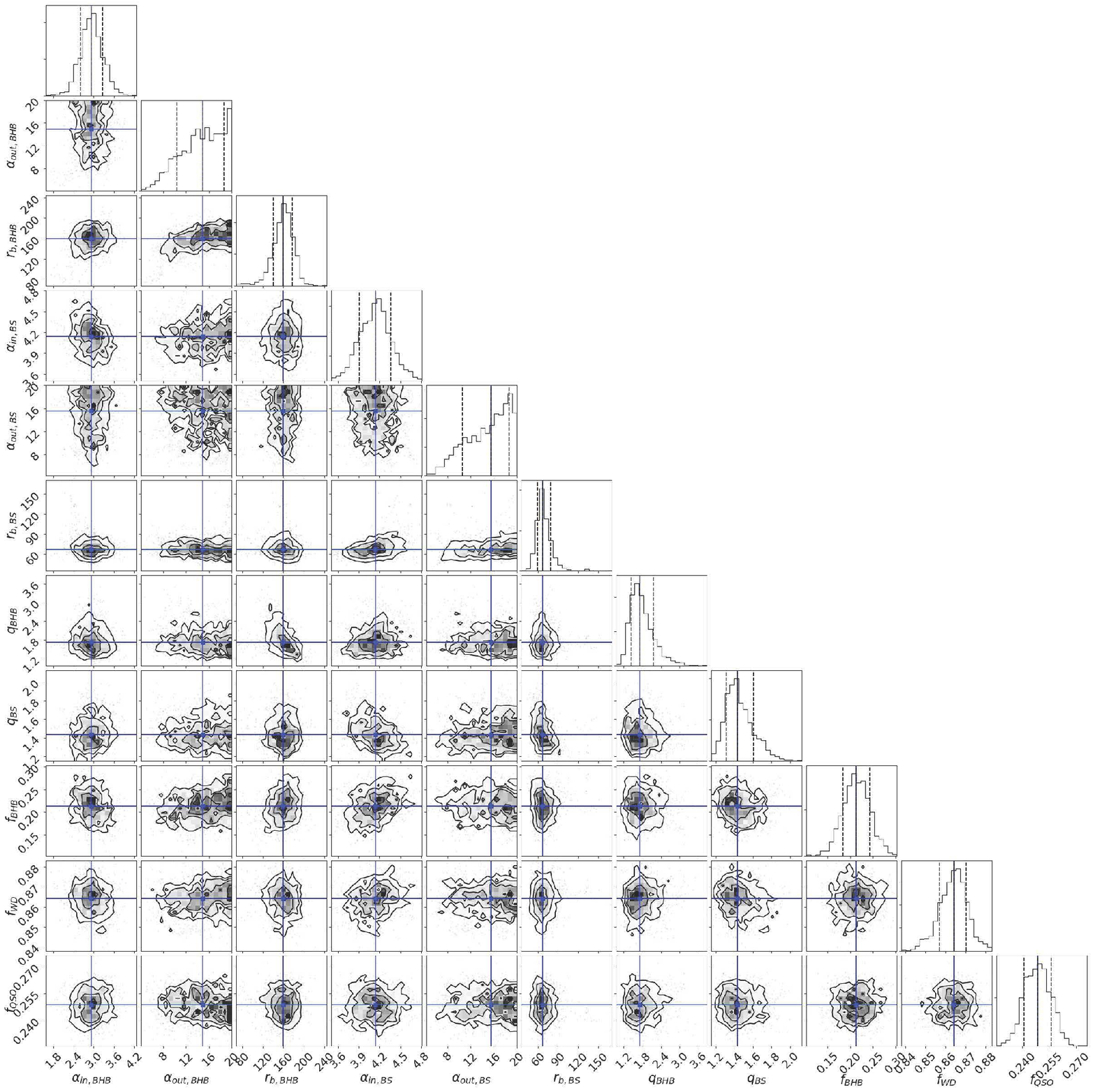}
\end{center}
\hspace*{10mm}
\caption{
MCMC results for ASPL (left) and ABPL (right panel).
}
\label{fig: Axisymmetric}
\end{figure*}

\begin{figure}[t!]
\begin{center}
\includegraphics[width=80mm]{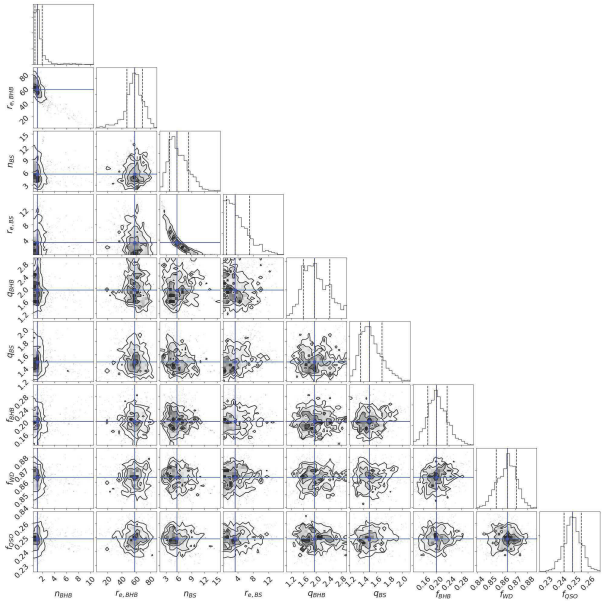}
\end{center}
\hspace*{10mm}
\caption{
MCMC results for Einasto model.
}
\label{fig: Einasto}
\end{figure}

Table \ref{tab:best_fit} shows the best fit parameters for the models of SSPL, SBPL, 
ASPL, ABPL and Einasto density profiles, respectively. The difference in the BIC values
relative to that for the best fit case (ABPL) is also listed in 
the last column. Figures \ref{fig: Spherical}, \ref{fig: Axisymmetric} and \ref{fig: Einasto}
 show the MCMC results for these models. We note that as given 
in Equation (\ref{eq: sampleslection}), these results correspond to the sample with the 
magnitude range of $18.5 < g < 23.5$, suggesting BHBs at about $r = 36 \sim 360$~kpc and 
BSs at about $r = 16 \sim 160$~kpc. The main properties of the results are summarized as follows.
\begin{itemize}
\item Both single power-law models of SSPL and ASPL reveal similar index values, i.e., 
BHBs are fit to $\alpha = 3.7 \sim 3.8$, whereas BSs show steeper density profiles of 
$\alpha = 4.4 \sim 4.6$.
\item For BHBs, double power-law models (SBPL and ABPL) show slightly shallower 
profiles at $r < r_{\rm b}$ than the corresponding single power-law models (SSPL and 
ASPL) expressed as $\alpha_{\rm in} < \alpha$. For BSs, $\alpha_{\rm in}$ is 
basically the same as $\alpha$ within the 1$\sigma$ error.
\item The non-spherical models of ASPL and ASBL suggest a prolate shape of $q = 1.4 
\sim 1.8$.
\item Both double power-law models of SBPL and ABPL show very steep index values 
of $\alpha_{\rm out}$ for both BHBs and BSs, suggesting outer boundaries in both 
populations.
\item ABPL provides the lowest BIC, thus is most likely among the given models.
\item The best-fit parameters for calculating the fractions of the populations,
$f_{\rm BHB}$, $f_{\rm WD}$ and $f_{\rm QSO}$ are basically the same for different models.
We then obtain the fraction of each population as
$\tilde{f}_{\rm BHB}=0.0195-0.0218$,
$\tilde{f}_{\rm BS} =0.0781-0.0815$ and
$\tilde{f}_{\rm WD}=0.649-0.658$.
\end{itemize}


We also consider the effects of some modification for the parameters of WDs,
especially the scale height, $H_2$, for the thick-disk component, which is generally
uncertain. We examine the case when the value of $H_2$ is modified from $0.9$~kpc
to $2$~kpc for ABPL. It is found that the change in $\alpha_{\rm in}$
is confined to be about 10\%. The changes in $\alpha_{\rm out}$ and $r_b$ are in the
range of 13 to 21\%, whereas the change in $q$ is up to 55\%, although the halo shape
remains to be prolate. Thus, we conclude that some minor modification for the
parameters of WDs do not affect the general properties of the density profile for
both BHBs and BSs.

\subsection{Comparison with our previous work}

In \citet{Fukushima2018}, we reported our work based on the simple color cuts in 
$griz$ band for the selection of BHBs using the S16A data of HSC-SSP over $\sim 
300$~deg$^2$ area. The main results in that paper for the case excluding the fields 
containing known substructures are roughly the same as those presented here, although 
there are some detailed differences. These previous results are summarized as $\alpha \simeq 
3.5$ and $q \simeq 1.3$ for ASPL and $\alpha_{\rm in} \simeq 3.2$, $\alpha_{\rm out} 
\simeq 5.3$, $q \simeq 1.5$ and $r_{\rm eff} \simeq 210$~kpc for ABPL. This suggests that 
compared with these previous results, the current analysis gives somewhat steep 
$\alpha$ and large $q$ for ASPL, whereas $\alpha_{\rm out}$ is made quite steep for 
ABPL. This may be caused by the removal of more BS contamination from candidate 
BHBs in the outskirts of the halo based on the current Bayesian analysis than those 
made in our previous work, as well as the use of the HSC-SSP data over much larger 
survey areas.

To assess the above statement, we analyze the HSC-SSP data adopted in 
\citet{Fukushima2018} (with a magnitude limit of $g < 22.5$) but using the method 
developed here. We obtain, for BHBs, $\alpha = 4.12^{+0.83}_{-0.60}$ and $q = 
1.08^{+1.09}_{-0.55}$ for ASPL and $\alpha_{\rm in} =4.00^{+0.81}_{-0.89}$, 
$\alpha_{\rm out} = 9.80^{+6.67}_{-4.99}$, $q = 1.00^{+1.65}_{-0.51}$ and $r_{\rm eff} 
\sim 158.9^{+59.5}_{-61.9}$~kpc for ABPL. Thus, due to the removal of more BS contamination
in the outskirts of the halo, the current new analysis leads somewhat steeper $\alpha$,
although this change remains within the 1$\sigma$ error. In the current work using the S18A data, 
the axial ratio, $q$, is made larger than that using the S16A data. This may be due to 
the increase of the S18A sample at high Galactic latitudes, where the sensitively to the 
prolate shape of the stellar halo can be increased. In this manner, it is possible to 
understand the changes in the results from our previous work, and the current 
work is expected to provide more realistic model parameters having smaller errors.

\begin{figure*}[t!]
\begin{center}
\includegraphics[width=80mm]{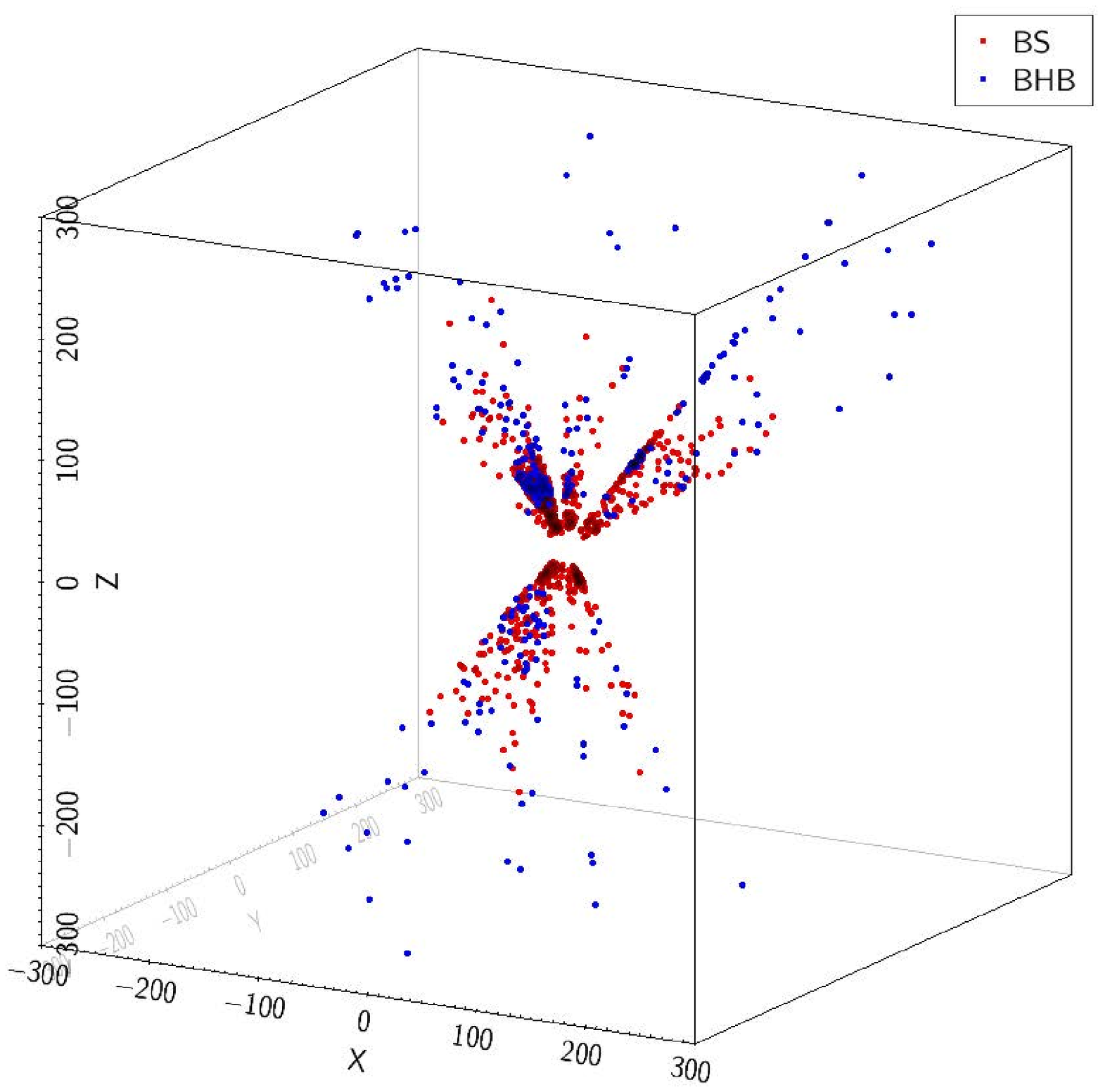}
\includegraphics[width=80mm]{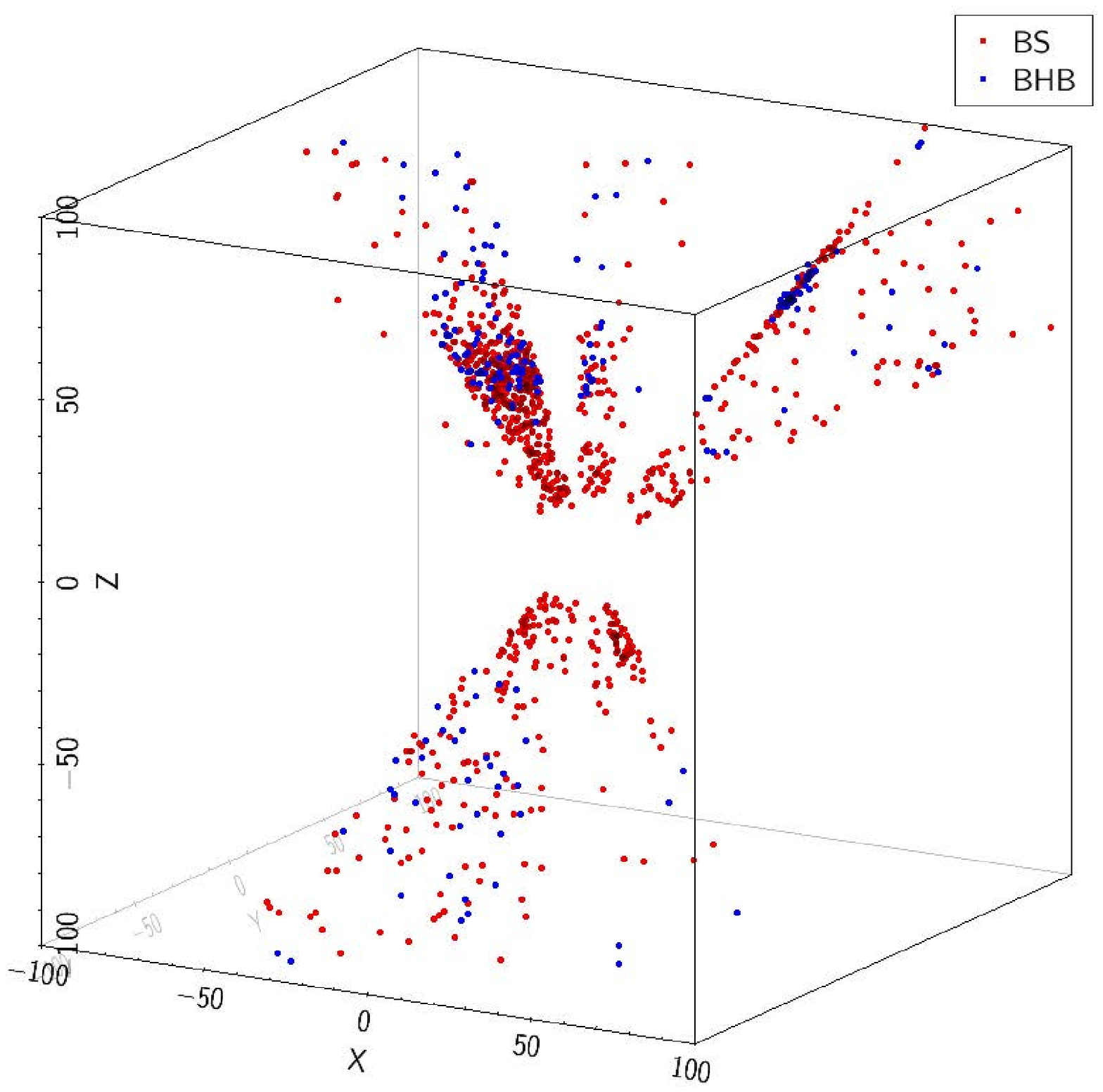}
\end{center}
\hspace*{10mm}
\caption{
Three-dimensional distributions of BHBs (blue points) and BSs (red points) selected from
those having high probabilities as BHBs [$p({\rm BHB}|x)>0.7$] and BSs [$p({\rm BS}|x)>0.7$],
respectively, as defined in Equation (\ref{eq: BHB_prob}). The left panel shows the box over
$-200 \le x,y,z \le 200$~kpc and the right panel shows the zoom-in view of the inner region
over $-100 \le x,y,z \le 100$~kpc.
}
\label{fig: 3d_map}
\end{figure*}

\subsection{Three-dimensional maps of BHBs and BSs}

So far, we focus on the smooth parts of the stellar halo by excluding the fields,
GAMA15H and XMM-LSS, which contain the known substructures including the Sgr stream.
Given that the parameters $f_{\rm BHB}$, $f_{\rm WD}$ and $f_{\rm QSO}$ basically remain
the same among different density models, it is possible to derive the probability that
a given target is either of a BHB, BS, WD, QSO or galaxy. For instance, the probability
of a BHB is given as
\begin{equation}
p({\rm BHB}|x) = \frac{p(x|{\rm BHB})f_{\rm BHB}}{ \sum_{i=1}^{4}p(x|A_i) \tilde{f}_i 
                                + p(x|{\rm galaxy})\frac{1-P_{\rm star}}{P_{\rm star}} } \ ,
\label{eq: BHB_prob}
\end{equation}
where $x$ shows each sample and $i$ denotes a component (BHB, BS, WD and QSO).

Figure \ref{fig: 3d_map} shows the three-dimensional maps for the sample with $p({\rm BHB}|x)$
larger than 70\% (blue points) and $p({\rm BS}|x)$ larger than 70\% (red points)
using all the survey fields.
There is a substructure associated with the Sgr stream at around $(x,y,z) = (-20, 10, 40)$~kpc
as seen for both BHBs and BSs. Sextans dSph is visible at $(x,y,z) = (40, 60, 60)$~kpc, and
there appears an overdensity at $(x,y,z) = (0, -40, -50)$~kpc, which might be the tidal debris
from the Large Magellanic Cloud \citep{Diaz2012}.

\begin{figure}[t!]
\begin{center}
\includegraphics[width=80mm]{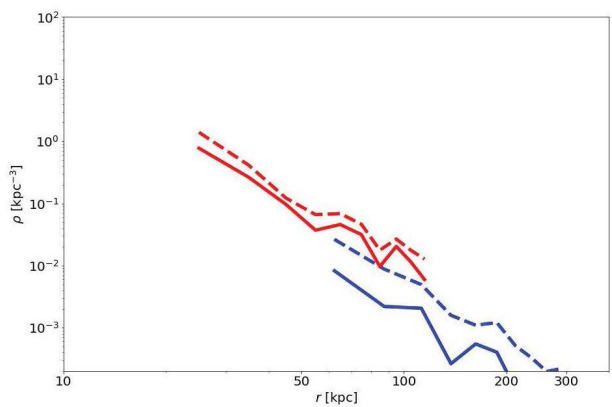}
\end{center}
\hspace*{10mm}
\caption{
The density distribution of BHBs (blue lines) and BSs (red lines),
where the solid (dashed) lines correspond to these stars having probabilities
larger than 80\% (70\%), namely $p({\rm BHB}|x)>0.8$ and $p({\rm BS}|x)>0.8$
($p({\rm BHB}|x)>0.7$ and $p({\rm BS}|x)>0.7$).
}
\label{fig: binned_density}
\end{figure}

Figure \ref{fig: binned_density} shows the density distribution of BHBs (blue lines) and
BSs (red lines), where the solid (dashed) lines correspond to these stars having probabilities
larger than 80\% (70\%), namely $p({\rm BHB}|x)>0.8$ and $p({\rm BS}|x)>0.8$
($p({\rm BHB}|x)>0.7$ and $p({\rm BS}|x)>0.7$). It follows that these high-probability sample
stars show a signature of broken density profiles changed at $r \sim 160$~kpc for BHBs and
$r \sim 70$~kpc for BSs, respectively, as suggested from the best-fit models in the previous
subsection. We note that the actual density profiles are obtained over the integral of
these probability distributions in our Bayesinan method.

\begin{figure*}[t!]
\begin{center}
\includegraphics[width=160mm]{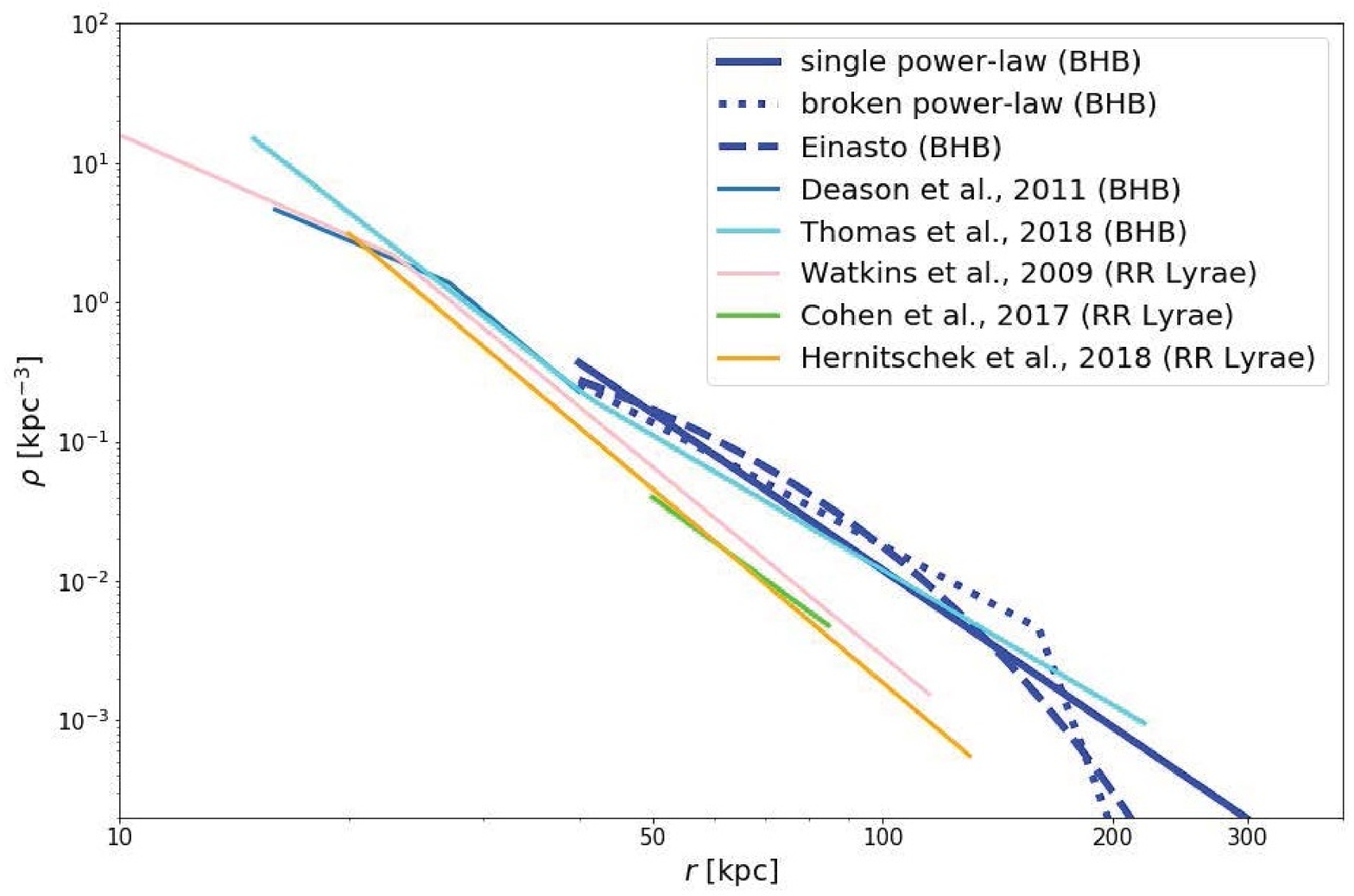}
\end{center}
\hspace*{10mm}
\caption{
Comparison of our best-fit models, the single power-law (blue solid line), broken power-law
(blue dotted line), and the Einasto profile (blue dashed line),
with other works using BHBs \citep{Deason2011,Thomas2018}
and RR Lyrae \citep{Watkins2009,Cohen2017,Hernitschek2018}.
}
\label{fig: comparison}
\end{figure*}

\section{Discussion}

\subsection{Comparison with other survey results}

Many previous surveys for tracing the MW stellar halo have been made, as mentioned 
in Section 1, but except for the following recent works, the most of the other surveys are 
devoted to the halo regions at Galactocentric radii well below $r = 100$~kpc.
In this subsection, we compare our results with the other surveys for $r$ as large as 100~kpc, which are
summarized in Figure \ref{fig: comparison}.

\citet{Thomas2018} recently combined their CFIS survey made in deep $u$-band with 
$griz$-band data from Pan-STARRS~1 to select candidate BHBs. Their analysis 
revealed that a broken power-law model with an inner/outer slope of $4.24/3.21$ at a 
break radius of 41.4 kpc is the best fitting case out to $r \sim 220$~kpc. This outer 
slope is similar to the inner slope of $\simeq 2.92$ in our ABPL model at $r < r_{\rm b} 
\simeq 160$~kpc, thus giving an approximate agreement. In contrast, their model of a 
fixed axial ratio showed $q \simeq 0.86$, i.e., an oblate halo. However, their alternative 
model allowing a varying $q$ suggests a prolate halo in the outer halo, which is 
consistent with our results.

The surveys using RRLs at $r$ as large as 100~kpc tend to provide different density 
slopes \citep{Watkins2009,Cohen2017,Hernitschek2018}. These works show $\alpha = 
4.0 \sim 4.5$ at $r > 25$~kpc, which is systematically steeper than the slopes obtained 
here for BHBs, but consistent with those for BSs located at similar radii to RRLs 
($\alpha \simeq 4.50$ for ASPL, $\alpha_{\rm out} \simeq 4.22$ for ABPL). This 
implies that the difference in the value of the density slope for BHBs from that for RRLs 
is due to the difference in the range of Galactocentric radii for the adopted sample. 
Another possible reason for the different slopes may be due to the intrinsically different 
radial distribution for a different stellar sample, depending on the formation history of a 
stellar halo associated with merging/accretion of progenitor dwarf galaxies as 
discussed in the next subsection.

Our current work suggests that the density slope of the MW halo is somewhat shallower 
at $r > 100$~kpc as probed by BHBs than the slope at radii near and below $\sim 
100$~kpc. Also, the very steep slope at radii above $\simeq 160$~kpc for BHBs may 
suggest a sharp outer edge of the stellar halo. On the other hand, a steeper 
$\alpha$ and smaller break radius ($r_{\rm b} \simeq 70$~kpc) for BSs may be due to 
the intrinsically more centrally concentrated spatial distribution of BSs than BHBs in 
the MW halo. This may be caused by the more centrally distributed BSs in progenitor 
dwarf galaxies (e.g., \cite{Wang2018}): in the course of merging/accretion of dwarf galaxies,
these denser, central parts can fall into the more central parts of the MW halo due to the effects of 
dynamical friction, so that the debris after the destruction of dwarf galaxies reflect the 
original internal distribution inside dwarf galaxies.

\subsection{Possible constraints on the past accretion history}

To infer what constraints from the current analysis of BHBs can be made on the past 
accretion history of the MW halo, we compare our results with the suite of hydrodynamical simulations for
galaxy formation by \citet{Rodriguez-Gomez2016} using the Illustris Project \citep{Genel2014,
Vogelsberger2014a,Vogelsberger2014b}. \citet{Rodriguez-Gomez2016} investigated the formation
of galaxies over a wide range of stellar masses, $M_{\ast}=10^{9}-10^{12} M_{\odot}$, and
obtained the relative contribution of the so-called {\it in situ.} halo (main progeitor halo)
with respect to the {\it ex situ.} halo (accreted stellar system from outside) component.
It is found that these halo components are spatially segregated, with {\it in situ.} halo
dominating the innermost regions of the halo space, and {\it ex situ.} halo being deposited
at larger Galactocentric distances in order of decreasing merger mass ratio.
These properties are well summarized in their Figure 10: the {\it in situ.} component
shows a steep density profile below the transition radius, whereas the {\it ex situ.} component
beyond this radius provides a shallow slope having an outer boundary.
This theoretical prediction may well reproduce the change of the halo density profile
mentioned in the previous subsection, namely the steep profile in the inner halo probed
by RRLs, which were possibly formed {\it in situ.}, and the shallow profile in the outer halo
reported here using BHBs, which were originated from the {\it ex situ.} component.

\section{Conclusions}

Using the HSC-SSP Wide layer data obtained until 2018 April (S18A), which covers
$\sim 550$~deg$^2$ area, we have selected candidate BHB stars based on 
the $(g,r,i,z)$ photometry, where $z$-band brightness can be used to probe a surface 
gravity of a BHB star against other A-type stars. In contrast to our previous work 
reported in \citet{Fukushima2018}, where the simple color cuts were adopted for the 
selection of BHBs, we have developed an extensive Bayesian method to minimize the 
effects of non-BHB contamination as much as possible.
In this analysis, the distributions of the template BHBs and non-BHB populations are
represented as a mixture of multiple Gaussians in the color-color diagrams defined in $griz$ band.
This method is especially effective for removing BS contamination in a statistically
significant manner.

Applying to the sample with $18.5 < g < 23.5$, which, for candidate BHBs, correspond 
to the positions of Galactocentric radii at $r = 36 \sim 360$~kpc, we have obtained the 
density slopes of BHBs for a single power-law model as $\alpha = 
3.74^{+0.21}_{-0.22}$ and for a broken power-law model as $\alpha_{\rm 
in}=2.92^{+0.33}_{-0.33}$ and $\alpha_{\rm out}=15.0^{+3.7}_{-4.5}$ divided at a 
radius of $r_{\rm b}=160^{+18}_{-19}$~kpc. The latter power-law model appears most 
likely according to BIC. For the models allowing a non-spherical halo 
shape, an axial ratio of $q = 1.72^{+0.44}_{-0.28}$ corresponding to a prolate shape
is the most likely case. It is also suggested from a very steep $\alpha_{\rm out}$ that
the MW stellar halo may have a sharp boundary at $r = r_{\rm b} \simeq 160$~kpc,
although this needs to be assessed using the further survey data.

The density slope obtained in this work is basically in agreement with that from the 
recent CFIS survey for BHBs \citep{Thomas2018}. However, it is systematically 
shallower than the slope derived from RRL stars at $r$ below $\sim 100$~kpc 
\citep{Cohen2017, Hernitschek2018}. This may be simply due to the different radial 
range of each sample, $r < 100$~kpc for RRLs and $50 < r < 360$~kpc for BHBs, or 
RRLs may have an intrinsically more centrally concentrated distribution than BHBs.
However, before concluding so, we require much larger data for BHBs
obtained by the completion of the HSC-SSP survey with a goal of $\sim 
1,400$~deg$^2$. Also, to interpret such observational results in the form of the past 
merging history, more extensive numerical simulations for the formation of stellar halos 
will be important, where not only accretion/merging of satellites from outside but also 
the {\it in situ.} formation of halo stars are properly taken into account.

\begin{ack}
This work is supported in part by JSPS Grant-in-Aid for Scientific 
Research (B) (No. 25287062) and MEXT Grant-in-Aid for Scientific Research
(No. 16H01086, 17H01101 and 18H04334 for MC,
No. 18H04359 and No. 18J00277 for KH).
TM is supported by Grant-in-Aid for JSPS Fellows (No. 18J11326).

The Hyper Suprime-Cam (HSC) collaboration includes the astronomical
communities of Japan and Taiwan, and Princeton University.  The HSC
instrumentation and software were developed by the National
Astronomical Observatory of Japan (NAOJ), the Kavli Institute for the
Physics and Mathematics of the Universe (Kavli IPMU), the University
of Tokyo, the High Energy Accelerator Research Organization (KEK), the
Academia Sinica Institute for Astronomy and Astrophysics in Taiwan
(ASIAA), and Princeton University.  Funding was contributed by the FIRST 
program from Japanese Cabinet Office, the Ministry of Education, Culture, 
Sports, Science and Technology (MEXT), the Japan Society for the 
Promotion of Science (JSPS),  Japan Science and Technology Agency 
(JST),  the Toray Science  Foundation, NAOJ, Kavli IPMU, KEK, ASIAA,  
and Princeton University.
This paper makes use of software developed for the Large Synoptic Survey Telescope. We thank the
LSST Project for making their code freely available. The Pan-STARRS1 (PS1) Surveys have been made
possible through contributions of the Institute for Astronomy, the University of Hawaii,
the Pan-STARRS Project Office, the Max-Planck Society and its participating institutes, the Max Planck Institute for
Astronomy and the Max Planck Institute for Extraterrestrial Physics, The Johns Hopkins University,
Durham University, the University of Edinburgh, Queen's University Belfast, the Harvard-Smithsonian
Center for Astrophysics, the Las Cumbres Observatory Global Telescope Network Incorporated, the
National Central University of Taiwan, the Space Telescope Science Institute, the National Aeronautics
and Space Administration under Grant No. NNX08AR22G issued through the Planetary Science Division
of the NASA Science Mission Directorate, the National Science Foundation under Grant
No.AST-1238877, the University of Maryland, and Eotvos Lorand University (ELTE).
\end{ack}



\begin{thebibliography}{}
\bibitem[Abazajian et al.(2004)]{Abazajian2004} Abazajian, K., Adelman-McCarthy, J. K.,
Ag\"ueros, M. A., et al. 2004, \aj, 128, 502

\bibitem[Aihara et al.(2018a)]{Aihara2018a} Aihara, H., Arimoto, N., Bickerton, S.,
et al. 2018a, \pasj, 70, S4 (arXiv:1704.05858)

\bibitem[Aihara et al.(2018b)]{Aihara2018b} Aihara, H., Armstrong, R., Bickerton, S.,
et al. 2018b, \pasj, 70, S8 (arXiv:1702.08449)

\bibitem[Axelrod et al.(2010)]{Axelrod2010} Axelrod, T., Kantor, J., Lupton, R.~H.,
\& Pierfederici, F. 2010, in   \procspie, Vol. 7740, Software and Cyberinfrastructure
for Astronomy, 774015

\bibitem[Belokurov et al.(2006)]{Belokurov2006} Belokurov, V., Zucker, D. B., Evans, N. W.,
et al. 2006, \apjl, 647, L11

\bibitem[Bland-Hawthorn \& Freeman(2014)]{Bland-Hawthorn2014} Bland-Hawthorn, J.,
\& Freeman, K., 2014, The Origin of the Galaxy and Local Group, Saas-Fee Advanced Course,
Vol. 37 (Springer)

\bibitem[Bosch et al.(2018)]{Bosch2018} Bosch, J., Armstrong, R., Bickerton, S.,
 et al. 2018, \pasj, 70, S5

\bibitem[Bovy et al.(2011)]{Bovy2011} Bovy J., Hogg D. W., Roweis S. T., 2011, Annals of Applied Statistics, 5, 1657

\bibitem[Bullock \& Johnston(2005)]{Bullock2005} Bullock, J. S. \& Johnston, K. V. 2005,
\apj, 635, 931

\bibitem[Chen et al.(2001)]{Chen2001} Chen, B., Stoughton, C., Smith, J. A., et al. 2001,
\apj, 553, 184

\bibitem[Cohen et al.(2015)]{Cohen2015} Cohen, J. G., Sesar, B., Banholzer, S., the PTF
Collaboration, 2015, IAU General Assembly, 22, 2255152

\bibitem[Cohen et al.(2017)]{Cohen2017}  Cohen, J. G., Sesar, B., Banholzer, S. et al. 2017, \apj, 849, 150

\bibitem[Cooper et al.(2010)]{Cooper2010} Cooper, A. P. et al., 2010, \mnras, 406. 744

\bibitem[Deason et al.(2011)]{Deason2011} Deason, A. J., Belokurov, V., \& Evans, N. W.
2011, \mnras, 416, 2903

\bibitem[Deason et al.(2014)]{Deason2014} Deason, A. J., Belokurov, V., Koposov S. E., Rockosi C. M. 2014, \apj, 787, 30

\bibitem[Deason et al.(2018a)]{Deason2018a} Deason, A. J., Belokurov, V., Koposov S. E. 2018a,
\apj, 852, 118

\bibitem[Deason et al.(2018b)]{Deason2018b} Deason, A. J., Belokurov, V., Koposov S. E., \&
Lancaster, L. 2018b, arXiv:1805.10288

\bibitem[Diaz \& Bekki(2012)]{Diaz2012} Diaz, J., \& Bekki, K. 2012, \apj, 750, 36

\bibitem[Einasto(1965)]{Einasto1965} Einasto, J. 1965, Trudy Inst. Astroz. Alma-Ata, 5, 87

\bibitem[Feltzing \& Chiba(2013)]{Feltzing2013} Feltzing, S., \& Chiba, M. 2013,
New Astronomy Reviews, 57, 80

\bibitem[Foreman-Mackey et al.(2013)]{Foreman-Mackey2013} Foreman-Mackey D., 
Hogg D. W., Lang D., Goodman J., 2013, PASP, 125, 306

\bibitem[Fukushima et al.(2018)]{Fukushima2018} Fukushima, T., Chiba, M., et al. 2018,
\pasj, 70, 69

\bibitem[Furusawa et al.(2018)]{Furusawa2018} Furusawa, H., et al. 2018, \pasj, 70, S3

\bibitem[Genel et al.(2014)]{Genel2014} Genel, S. et al. 2014, \mnras, 445, 175

\bibitem[Goodman \& Weare(2010)]{Goodman2010} Goodman J., Weare J., 2010,
Comm. App. Math. Comp. Sci., 5, 65

\bibitem[Graham et al.(2006)]{Graham2006} Graham, A. W., Merritt, D., Moore, B., et al.
2006, AJ, 132, 2701

\bibitem[Grand et al.(2017)]{Grand2017} Grand, R. J. J., et al. 2017, \mnras, 467, 179

\bibitem[Gunn \& Stryker(1983)]{Gunn1983} Gunn, J. E., \& Stryker, L. L., 1983,
\apjs, 52, 121

\bibitem[Helmi \& White(1999)]{Helmi1999} Helmi, A. \& White, S. D. M., 1999, \mnras, 307, 495

\bibitem[Helmi(2008)]{Helmi2008} Helmi, A. 2008, A\&ARV, 15, 145

\bibitem[Hernitschek et al.(2018)]{Hernitschek2018} Hernitschek, N., Cohe, J. G., Rix, H.-W.,
et al., 2018, \apj, 859, 31

\bibitem[Holoien et al.(2017)]{Holoien2017} Holoien T. W.-S., Marshall P. J., Wechsler R. H., 2017, \aj, 153, 249

\bibitem[Homma et al.(2016)]{Homma2016} Homma, D., Chiba, M., Okamoto, S., et al.
2016, \apj, 832, 21

\bibitem[Ibata et al.(1995)]{Ibata1995}Ibata, R. A., Gilmore, G., Irwin, M. J.
1995, \mnras, 277, 781

\bibitem[Ivezi\'c et al.(2008)]{Ivezic2008} Ivezi\'c, Z., Axelrod, T., Brandt, W. N., et al.
2008, \aj, 176, 1

\bibitem[Ivezi\'c, Beers \& Juric(2012)]{Ivezic2012} Ivezi\'c, Z., Beers, T. C., \& Juri\'c, M.
2012, ARAA, 50, 251

\bibitem[Juri\'c et al.(2008)]{Juric2008} Juri\'c, M., Ivezi\'c, Z., Brooks, A., et al.
2008, \apj, 673, 864

\bibitem[Juri\'c et al.(2017)]{Juric2017} Juri\'c, M., Kantor, J., Lim, K.-T., et al.
2017, ASPC, 512, 279

\bibitem[Kawanomoto et al.(2018)]{Kawanomoto2018} Kawanomoto, S. et al. 2018, \pasj, 70, 66

\bibitem[Keller et al.(2008)]{Keller2008} Keller, S. C., Murphy, S., Prior, S., Da Costa, G., \& Schmidt, B. 2008, \apj, 678, 851

\bibitem[Kepler et al.(2015)]{Kepler2015} Kepler, S.~O., Pelisoli, I., Koester, D., et al.
2015, \mnras, 446, 4078

\bibitem[Kepler et al.(2016)]{Kepler2016} Kepler, S.~O., Pelisoli, I., Koester, D., et al.
2016, \mnras, 455, 3413

\bibitem[Kleinman et al.(2013)]{Kleinman2013} Kleinman, S.~J., Kepler, S.~O., Koester, D.,
et al. 2013, \apjs, 204, 5

\bibitem[Komiyama et al.(2018)]{Komiyama2018} Komiyama, Y., et al. 2018, \pasj, 70, S2

\bibitem[Lee et al.(2008)]{Lee2008} Lee, Y.~S., Beers, T.~C., Sivarani, T., et al.
2008, \aj, 136, 2022.

\bibitem[Lenz et al.(1998)]{Lenz1998} Lenz, D. D., Newberg, J., Rosner, R., et al.
1998, \apjs, 119, 121

\bibitem[Magnier et al.(2013)]{Magnier2013} Magnier, E. A., Schlafly, E., Finkbeiner, D.,
et al. 2013, \apjs, 205, 20

\bibitem[Miyazaki et al.(2018)]{Miyazaki2018} Miyazaki, S., Komiyama, Y., Kawanomoto, S.
et al. 2018, \pasj, 70, S1

\bibitem[Monachesi et al.(2018)]{Monachesi2018} Monachesi, A. et al. 2018, arXiv:1804.07798

\bibitem[Newberg \& Yanny(2005)]{Newberg2005} Newberg, H. J., \& Yanny, B. 2005,
JPhCS, 47, 195

\bibitem[P\^aris et al.(2018)]{Paris2018} P\^aris I. et al. 2018, \aap, 613, A51

\bibitem[Pillepich et al.(2014)]{Pillepich2014} Pillepich, A., Vogelsberger, M., Deason, A.,
et al. 2014, \mnras, 244, 237

\bibitem[Rodriguez-Gomez et al.(2016)]{Rodriguez-Gomez2016} Rodriguez-Gomez, V.,
Pillepich, A., Sales, L. V. et al. 2016, \mnras, 458, 2371

\bibitem[Schlafly \& Finkbeiner(2011)]{Schlafly2011} Schlafly, E. F., \& Finkbeiner, D. P.
2011, \apj, 737, 103

\bibitem[Schlafly et al. (2012)]{Schlafly2012} Schlafly, E. F., Finkbeiner, D. P.,
Juri\'c, M., et al. 2012, \apj, 756, 158

\bibitem[Searle \& Zinn(1978)]{Searle1978} Searle, L., \& Zinn, R. 1978, \apj, 225, 357

\bibitem[Sesar et al.(2011)]{Sesar2011} Sesar, B, Juri\'c, M., \& Ivezi\'c, Z.
2011, \apj, 731, 4

\bibitem[Sirko et al.(2004)]{Sirko2004} Sirko, E., Goodman, J., Knapp, G. R., et al.
2004, \aj, 127, 899

\bibitem[Slater et al.(2016)]{Slater2016} Slater, C. T., Nidever, D. L., Munn, J. A.,
Bell, E. F., \& Majewski, S. R. 2016, \apj, 832, 206

\bibitem[Sluis \& Arnold(1998)]{Sluis1998} Sluis, A. P. N., \& 
Arnold, R. A. 1998, \mnras, 297, 732

\bibitem[Thomas et al.(2018)]{Thomas2018} Thomas, G. A., McConnachie, A. W.,
Ibata, R. A., et al. 2018, \mnras, 481, 5223

\bibitem[Tonry et al.(2012)]{Tonry2012} Tonry, J. L., Stubbs, C. W., Lykke, K. R.,
et al. 2012, \apj, 750, 99

\bibitem[Vickers et al.(2012)]{Vickers2012} Vickers, J. J., Grebel, E. K., \& Huxor, A. P.
2012, \aj, 143, 86

\bibitem[Vivas et al.(2016)]{Vivas2016} Vivas, A. K., Zinn, R., Farmer, J., et al.
2016, \apj, in press (arXiv:1608.08981)

\bibitem[Vogelsberger et al.(2014a)]{Vogelsberger2014a} Vogelsberger, M. et al. 2014a,
\nat, 509, 177

\bibitem[Vogelsberger et al.(2014b)]{Vogelsberger2014b} Vogelsberger, M. et al. 2014b,
\mnras, 444, 1518

\bibitem[Wang et al.(2018)]{Wang2018} Wang, M.-Y. et al. 2018, preprint (arXiv: 1809.07801)

\bibitem[Watkins et al.(2009)]{Watkins2009} Watkins, L. L. et al. 2009, \mnras, 398, 1757

\bibitem[Xu et al.(2018)]{Xu2018} Xu, Y., Liu, C., Xue, X.-X., et al. 2018, \mnras,
473, 1244

\bibitem[Xue et al.(2011)]{Xue2011} Xue, X.-X., Rix, H.-W., Yanny, B., et al. 2011,
\apj, 738, 79

\bibitem[Yanny et al.(2000)]{Yanny2000} Yanny, B., Newberf, H. J., Kent, S., et al.
2000, \apj, 540,825 


\end{thebibliography}
\end{document}